\documentclass[12pt]{article}

\newcommand{\blind}{0}

\usepackage{amsmath}
\usepackage{amssymb}
\usepackage{algorithm} 
\usepackage{algpseudocode}
\usepackage{graphicx,psfrag,epsf}
\usepackage{enumerate}

\usepackage{url} 
\usepackage{amsfonts}
\usepackage{graphicx} 
\usepackage[a4paper]{geometry}

\usepackage[sfdefault]{roboto}  
\usepackage{natbib}
\usepackage[english]{babel}
\usepackage{xcolor}
\usepackage{bm}

\usepackage{subcaption}

\newcommand{\matr}[1]{\mathbf{#1}}

\newcommand{\calN}{{\cal N}}

\newcommand{\calO}{{\cal O}}

\newcommand{\calD}{{\cal D}}

\newcommand{\norm}[1]{\left\lVert#1\right\rVert}

\newcommand{\Tr}{\mathrm{Tr}}

\renewcommand{\Vec}{\bm}
\newcommand{\vy}{\Vec{y}}
\newcommand{\vu}{{\Vec{u}}}
\newcommand{\vn}{{\Vec{n}}}

\newcommand{\vz}{{\Vec{z}}}
\newcommand{\vv}{{\Vec{v}}}

\newcommand{\vw}{{\Vec{w}}}

\newcommand{\vx}{\Vec{x}}
\newcommand{\vs}{\Vec{s}}
\newcommand{\vmu}{\Vec{\mu}}
\newcommand{\vbeta}{\Vec{\beta}}

\newcommand{\vomega}{\Vec{\omega}}

\newcommand{\vtheta}{\Vec{\theta}}
\newcommand{\vphi}{\Vec{\phi}}
\newcommand{\vepsilon}{\Vec{\epsilon}}
\newcommand{\vzero}{\Vec{0}}
\newcommand{\rT}{ {\top} }
\newcommand{\rS}{ {\mathrm{S}} }
\newcommand{\rR}{ {\text{Right}} }
\newcommand{\rL}{ {\text{Left}} }
\newcommand{\rU}{ {\text{Top}} }
\newcommand{\rD}{ {\text{Bottom}} }
\newcommand{\rC}{ { \mathrm{C}} }
\newcommand{\rV}{ {\mathrm{V}} }

\newcommand{\dt} {\Delta t}
\newcommand{\dA} {\mathrm{d}A}

\newcommand{\WM} {\text{WM}}
\begin{document}


\def\spacingset#1{\renewcommand{\baselinestretch}%
{#1}\small\normalsize} \spacingset{1}


\if0\blind
{
  \title{\Large Non-stationary Spatio-Temporal Modeling Using the Stochastic Advection-Diffusion Equation}
  
  \author{\normalsize Martin Outzen Berild\thanks{Corresponding author, moutzebe@gmail.com} and Geir-Arne Fuglstad \\
    \small Department of Mathematical Sciences,\\ \small Norwegian University of Science and Technology, Norway}
    \date{}
  \maketitle

} \fi
\if1\blind
{
  \bigskip
  \bigskip
  \bigskip
  \begin{center}
    {\LARGE\bf Non-stationary Spatio-Temporal Modeling Using the Stochastic Advection-Diffusion Equation}
\end{center}
  \medskip
} \fi


\begin{abstract}

We construct flexible spatio-temporal models through stochastic partial differential
equations (SPDEs) where both diffusion and advection can be spatially varying.
Computations are done through a Gaussian Markov random field approximation of the
solution of the SPDE, which is constructed through a finite volume method.
The new flexible non-separable model is compared to a flexible separable model
both for reconstruction and forecasting, and evaluated in terms of root mean square errors
and continuous rank probability scores.
A simulation study demonstrates that the non-separable model performs better when
the data is simulated from a non-separable model with diffusion and advection.
Further, we estimate surrogate models for emulating the output of a ocean model
in Trondheimsfjorden, Norway, and simulate observations of autonomous underwater vehicles.
The results show that the flexible non-separable model outperforms the flexible separable
model for real-time prediction of unobserved locations. 

\end{abstract}

\noindent%

\vfill

\newpage

\section{Introduction}
\label{sec:introduction}
Spatio-temporal data consist of observations that have associated both spatial locations
and time points. For example, temperature or precipitation measured at different weather stations,
or the concentration of pollutants in the ocean measured from ships.
Developing new statistical models for analyzing such data is an active area of
research \citep{porcu_30_2021}.
The overarching goal is first to accurately capture the complex interactions between
space and time in natural phenomena while also enabling efficient computation
for handling large datasets;
then, to leverage the estimated interactions to predict unobserved locations or to forecast.

A common approach to model spatio-temporal data is through Gaussian random fields (GRFs);
see, for example, \citet{cressie_statistics_2011}.
GRFs are fully characterized by a mean function and a covariance function,
which means that the key challenges are to model the mean structure and to
model the covariance structure.
However, the covariance function must be positive semi-definite, and this complicates the
construction of valid models with realistic features.

A simple way to guarantee a valid spatio-temporal covariance function is to take
the product of a spatial covariance function and a temporal covariance function.
This is called a \textit{separable} covariance function, but requires a very specific
and simple interaction between space and time, and the natural processes
advection and diffusion typically leads to non-separability.
Typically, one would construct separable covariance functions using \textit{stationary}
and \textit{isotropic} covariance functions such as the Matérn family of covariance
functions, which is popular due to the interpretable parameters and the flexibility in controlling 
the smoothness, range, and variance. 
The specification of covariance functions for highly dynamic and non-stationary processes
is known to be challenging \citep{cressie_statistics_2011,simpson_order_2012},
and assumptions such as isotropy, stationarity, and separability are often made to
improve computational tractability even though they are physically unrealistic.

Spatio-temporal datasets can be large, and direct computations are limited by a
complexity of $\calO(n^3)$, where $n$ is the number of observations.
This is often called the "big $n$ problem" \citep{banerjee_hierarchical_2003}. 
Many approaches have been suggested to handle large datasets \citep{heatonEtAl2019}.
Some include include low-rank matrices
\citep{cressie_fixed_2008,banerjee_gaussian_2008,wikle_low-rank_2010},
approximate likelihoods \citep{vecchia_estimation_1988,eidsvik_estimation_2014,katzfuss2021general},
and Gaussian Markov random fields (GMRFs) \citep{rue_gaussian_2005}. 
However, an issue with GMRFs is that they are defined
through Markov properties and conditional distributions, and are difficult to construct, in general.
\citet{lindgren_explicit_2011} proposed to overcome this issue by defining models using
stochastic partial differential equations (SPDEs) instead of covariance functions,
and then constructing GMRFs that approximate the solutions of the SPDEs.
 
SPDEs can be a powerful modeling tool for constructing non-separable spatio-temporal GRFs
as they can be related to partial differential equations that describe physical laws.
For instance, the stochastic heat or diffusion equation \citep{heine_models_1955,jones_models_1997,lindgren_diffusion-based_2023},
the stochastic wave equation \citep{carrizo_vergara_general_2022},
and the stochastic advection-diffusion equation \citep{liu_statistical_2022,clarotto_spde_2023}.
In the context of the SPDE approach, a common way to construct separable covariance
structures have been to use an autoregressive process of order 1 (AR(1) process),
where the temporal innovations are spatial GRFs \citep{rodriguez-iturbe_design_1974,cameletti_spatio-temporal_2013}.
Spatial non-stationarity can be introduced in the innovations using SPDEs with
spatially varying coefficients \citep{fuglstad_exploring_2014,fuglstad_does_2015,hildeman2021deformed,berild_spatially_2023}.
This gives a separable spatio-temporal covariance structure with spatial non-stationarity \citep{fuglstad_compression_2020}.

A promising approach to construct a class of non-separable covariance models with
spatial non-stationarity is to start with the stochastic advection-diffusion equation.
Previous work has approximated solutions through spectral representations \citep{sigrist_stochastic_2014,liu_statistical_2022},
or numerical methods such as finite element methods (FEMs) \citep{clarotto_spde_2023,lindgren_diffusion-based_2023}.
An issue with FEM is that it can be unstable in an advection-dominated setting \citep{clarotto_spde_2023}.
However, a more natural choice in a conservation law setting is to use the finite volume method (FVM) \citep{eymard2000finite},
as it is based on the flow of matter between cells in the discretization, and therefore, the FVM
is often closer to the physics of the problem \citep{leveque_finite_2002}.
The FVM has been used in \citet{fuglstad_exploring_2014,fuglstad_does_2015} and \citet{berild_spatially_2023}
in a purely spatial setting, but it has not been used
in spatio-temporal statistics even though it is prevalent in the field of computational fluid dynamics.

In this paper, we build on existing work on spatial non-stationarity \citep{fuglstad_does_2015,berild_spatially_2023} and stochastic advection-diffusion models \citep{clarotto_spde_2023} to develop more flexible models than previously considered. However, unlike \citet{berild_spatially_2023}, which uses three-dimensional space, we use two-dimensional space as computations would be too expensive with three-dimensional space and time unless spatial resolution were low.

The novel features of this work are: 1) parametrized spatially varying advection and anisotropic diffusion,
2) a non-stationary initial distribution,  3) the use of a FVM discretization of
the stochastic advection-diffusion equation, and 4) demonstrating the usefulness
of this approach to construct a statistical spatio-temporal surrogate model to a
deterministic numerical model. 
The new class of models allows for a wide range of spatio-temporal behavior to be captured.
With such a highly parameterized model, comes the challenge of navigating the likelihood
surface for such a high-dimensional parameter space.
We show that this is possible by borrowing optimization techniques from the machine
learning literature.

The behavior of the ocean is complex, and we need high-resolution
deterministic numerical models that are based on the Navier-Stokes equations to describe it well.
Such deterministic models are hard to update in real-time due to their computational complexity \citep{liu_statistical_2022}.  
In contrast, statistical models are highly effective for space-time interpolation
and prediction, making them suitable for real-time applications.
The challenge lies in sufficiently describing the complex spatio-temporal behavior.
We can take advantage of the complex numerical models to estimate complex statistical
surrogate models using the abundance of numerically simulated data.
Many authors have employed such an approach to estimate models used in ocean monitoring
with autonomous underwater vechicles (AUVs) but with some limitations and simplifications.
\citet{yaolin2023} estimated a purely spatial model through covariance functions
for ocean salinity field in 3D and used an adaptive sampling strategy
to choose locations for the AUVs to sample. \citet{foss_using_2022} used the output from
a numerical model to estimate an advection-diffusion model with covariate-based
advection and constant isotropic diffusion for monitoring mine tailings in the ocean.
\citet{berild_efficient_2024} estimated a non-stationary and anisotropic spatial
model through the SPDE approach for a 3D ocean salinity field. 

In Section~\ref{sec:theory}, we outline
the theoretical foundation for spatio-temporal modeling using SPDEs.
Then in Section~\ref{sec:method}, we introduce the proposed method for modeling spatio-temporal
data and elaborate on the parameterization of the stochastic advection-diffusion equation.
In Section~\ref{sec:examples}, we demonstrate the proposed method on a synthetic
dataset with a complex spatio-temporal structure and compare its predictive performance
to other spatio-temporal models. 
In Section~\ref{sec:applications}, we apply our method to emulate a numerical ocean
model, simulate an ocean monitoring scenario with AUVs, and compare the accuracy
of the proposed model to a non-stationary separable spatio-temporal model.
Finally, Section~\ref{sec:discussion} presents a discussion of the results and
our conclusions.
Some technical details such as the derivation of the numerical solution of the
stochastic advection-diffusion equation are deferred to the appendices.
The computer code used in this work is available at \url{https://github.com/berild/spdepy}.

\section{Spatio-temporal modeling with SPDEs}
\label{sec:theory}

\subsection{Non-stationary spatial modelling}
\label{subsec:spde}
The standard SPDE approach can be extended to non-stationary covariance structures
through spatially varying coefficients  \citep{lindgren_explicit_2011,lindgren_spde_2022}.
For a bounded domain of interest $\mathcal{D}\subset\mathbb{R}^2$,
we follow \citet{fuglstad_exploring_2014, fuglstad_does_2015} and consider the SPDE
\begin{equation}
    \label{eq:whittle_matern_spde}
    \left(\kappa(\boldsymbol{s})^2-\nabla\cdot\matr{H}(\boldsymbol{s})\nabla\right)u(\vs) = \mathcal{W}(\vs), \quad \vs \in\mathcal{D},
\end{equation}
where $\kappa(\cdot)$ is a positive function, $\nabla$ is the gradient, and $\matr{H}(\cdot)$ is a differentiable
spatially varying symmetric positive definite $2 \times 2$ matrix,
and $\mathcal{W}(\cdot)$ is Gaussian white noise. Note that $\mathcal{W}(\cdot)$ is a slight abuse of notation as Gaussian white noise must be viewed as a generalized Gaussian random field that does not make sense point-wise; see, e.g., \citet[Section 1.4.3]{adler2009random}. See \citet{bolin_rational_2020}
for more details on the technical requirements on $\kappa(\cdot)$ and $\matr{H}(\cdot)$.
We apply zero-flow Neumann boundary conditions, 
\[
    (\matr{H}(\vs)\nabla u(\vs))\cdot \boldsymbol{n}(\boldsymbol{s})=0, \quad \vs \in \partial \mathcal{D},
\]
where $\partial\mathcal{D}$ is the boundary of $\mathcal{D}$ and $\boldsymbol{n}(\cdot)$
is the outwards normal vector on the boundary.
This approach can be extended to $\mathbb{R}^3$ \citep{berild_spatially_2023} and
fractional powers can be used on the operator on the left-hand side to control
smoothness \citep{bolin_rational_2020}.
We write $u(\cdot)\sim \WM(\kappa(\cdot), \mathbf{H}(\cdot))$ to denote that $u(\cdot)$
has the Whittle-Matérn distribution arising from Equation \eqref{eq:whittle_matern_spde}
with the no-flow boundary condition, where the dependence on the domain $\mathcal{D}$
is suppressed in the notation.

Consider the stationary case that $\kappa(\cdot)\equiv\kappa_0$ and
$\mathbf{H}(\cdot)\equiv\mathbf{H}_0$, and ignore the boundary effects.
Then $u(\cdot)\sim \WM(\kappa_0, \mathbf{H}_0)$  leads to a Matérn covariance function
\begin{equation}
    \label{eq:whittle_matern_cov}
c(\vs, \vs') =
\sigma^2\left(\kappa_0\norm{\mathbf{H}_0^{-1/2}(\vs -
\vs')}\right)K_{1}\left(\kappa_0\norm{\mathbf{H}_0^{-1/2}(\vs -
\vs')}\right), \quad \vs, \vs'\in\mathbb{\mathcal{D}},
\end{equation}
where $\mathbf{H}_0^{-1/2}$ is the inverse of the square root of $\mathbf{H}_0$,
$\norm{\cdot}$ is the Euclidean norm, $\sigma^2$ is the marginal variance,
and $K_{1}$ is the modified Bessel function of the second kind, order $1$.
This is an anisotropic Matérn covariance function with smoothness $1$,
and as shown in \citet{fuglstad_exploring_2014}, the marginal variance is given by
\begin{equation*}
    \sigma^2 = 1/(4\pi\kappa_0^{2}\sqrt{\det \matr{H}_0}).
\end{equation*}
In practice, the influence of the boundary condition on an area of interest can be
reduced by extending the domain $\mathcal{D}$ to cover more than the area of interest
\citep{lindgren_explicit_2011}.


In the case that $\kappa(\cdot)$ and $\matr{H}(\cdot)$ are spatially varying,
there is no obvious way to avoid boundary effects since there is no unique way to
extend $\kappa(\cdot)$ and $\matr{H}(\cdot)$ to a larger domain than the area
of interest.
In practice, an option is to use a larger domain than the area of interest, and then
estimate $\kappa(\cdot)$ and $\matr{H}(\cdot)$ also outside the area of interest
\citep{fuglstad_does_2015}.
The anisotropic Laplacian in Equation \eqref{eq:whittle_matern_spde} induces
local anisotropy in the correlation at every location, and the combination of
$\kappa(\cdot)$ and $\matr{H}(\cdot)$ controls the spatially varying spatial
dependence and spatially varying marginal variance. See the discussion in
\citet{fuglstad_does_2015}.

In this paper, we will follow \citet{fuglstad_exploring_2014} and
parametrize $2\times 2$ symmetric positive-definite matrices through
\begin{equation}
\label{eq:change_basis}
\matr{H}(\vs) = \gamma(\vs) \matr{I}_2 + \vv(\vs)\vv(\vs)^\rT, \quad \vs\in\mathcal{D},
\end{equation}
where $\gamma(\cdot)>0 $ controls an isotropic baseline dependence, $\matr{I}$
is the identity matrix, and $\vv(\cdot)$ is a vector field that 
controls the direction and strength of extra anisotropy at each location.
For inference, we then expand $\log(\kappa(\cdot))$, $\log(\gamma(\cdot))$, $v_1(\cdot)$,
and $v_2(\cdot)$ in bases and estimate the coefficients of the basis functions
\citep{fuglstad_does_2015,berild_spatially_2023}. The details are discussed in Section \ref{subsec:parametrization}.

\subsection{Separable spatio-temporal modelling}
\label{subsec:separable}
A direct way to extend 
$\WM(\kappa(\cdot), \mathbf{H}(\cdot))$, to a spatio-temporal GRF is
to use a temporal AR(1) process, where the increments are $\WM(\kappa(\cdot), \mathbf{H}(\cdot))$. This gives 
\begin{equation}
    u^{k+1}(\vs) = a u^k(\vs) + \sqrt{1-a^2} \phi^k(\vs), \quad \vs\in\mathcal{D},
    \quad k = 1, \ldots, K-1,\label{eq:AR1-spde}
\end{equation}
where $K$ is the number of time steps, $a$, such that $|a|<1$, describes temporal autocorrelation,
and $u^1(\cdot), \phi^1(\cdot), \ldots,
\phi^{K-1}(\cdot)\overset{\text{iid}}{\sim}\WM(\kappa(\cdot), \mathbf{H}(\cdot))$.
\citet{fuglstad_compression_2020} used such a construction for a global model
with $\mathcal{D} = \mathbb{S}^2$, and \citet{cameletti_spatio-temporal_2013}
used this construction with $\kappa(\cdot)\equiv \kappa_0\sqrt{\tau}> 0$  and
$\matr{H}(\cdot)\equiv \tau\matr{I}$, where $\matr{I}$ is the $2\times 2$ identity matrix and $\tau > 0$ is a scaling parameter.

Introduce the GRF $\{B(\vs, t): \vs \in\mathcal{D}, t\in[0, \infty)\}$ such that
\begin{enumerate}
    \item $B(\cdot, 0) \equiv 0$,
    \item $B(\cdot, \cdot)$ has independent increments in time, and
    \item $(B(\cdot, t_2)-B(\cdot, t_1))/\sqrt{t_2-t_1} \sim
    \WM(\kappa(\cdot), \mathbf{H}(\cdot))$ for all $t_2>t_1\geq 0$.
\end{enumerate}
With an additional technical requirement that the sample paths of $B(\cdot, \cdot)$
are continuous in time, this is an extension of the traditional Brownian motion to
a spatio-temporal setting and can be formalized as a Q-Wiener process \citep[p. 81]{da2014stochastic}.

Following \citet{lindgren_spde_2022} and using the Q-Wiener process $B(\cdot, \cdot)$, one can view
the AR(1) process above as arising from an SPDE
\begin{equation}
    \frac{\partial}{\partial t}u(\vs,t)+c u(\vs,t) =
    \sqrt{2c}\frac{\partial}{\partial t}B(\vs,t), \quad \vs\in\mathcal{D},
    \quad t\in (0,\infty),
    \label{eq:sepSPDE}
\end{equation}
with the initial condition $u(\cdot, 0)\sim \WM(\kappa(\cdot), \mathbf{H}(\cdot))$.
Equation \eqref{eq:AR1-spde} can be viewed as a discrete approximation of the above SPDE,
and the parameter $a$ in Equation \eqref{eq:AR1-spde}
depends on $c$ and the temporal resolution used in the approximation.


\subsection{The stochastic advection-diffusion equation}
\label{subsec:advection_diffusion}
A weakness of separable spatio-temporal models is that they
cannot include non-separable phenomena such as diffusion and advection. 
Constructing non-separable spatio-temporal GRFs with spatially varying diffusion
and advection through covariance functions is hard, but they are easy to construct
through a stochastic advection-diffusion-reaction SPDE. We extend Equation \eqref{eq:sepSPDE} to
\begin{equation}
\label{eq:advection_diffusion}
\frac{\partial}{\partial t}u(\vs,t) + \mathrm{A}(\vs)u(\vs, t) = \tau \frac{\partial}{\partial t}B(\vs, t), \quad \vs\in\mathcal{D}, \quad t\in (0, \infty),
\end{equation}
where $\mathrm{A}(\cdot)$ is discussed in the next paragraph, $\tau > 0$ is
a scaling parameter, and the Q-Wiener process $B(\cdot, \cdot)$ has increments
described through $\WM(\kappa_\mathrm{F}(\cdot), \mathbf{H}_\mathrm{F}(\cdot))$.
The initial condition is $u(\cdot, 0)\sim \WM(\kappa_\mathrm{I}(\cdot), \mathbf{H}_\mathrm{I}(\cdot))$.
The subscripts ``F'' and ``I'' are short for ``Forcing'' and ``Initial''.
The boundary conditions are described in the next paragraph after defining the operator $\mathrm{A}(\cdot)$.

The advection, diffusion, and reaction are described as
\[
\mathrm{A}(\boldsymbol{s})u(\vs,\cdot) = \left(\kappa_{\mathrm{E}}(\vs)^2 - \nabla \cdot
\matr{H}_{\mathrm{E}}(\vs)\nabla\right)u(\vs,\cdot) +  \nabla\cdot(\vomega_\mathrm{E}(\vs) u(\vs,\cdot)), \quad \boldsymbol{s}\in\mathcal{D},
\]
where $\kappa_{\mathrm{E}}(\vs)^2u(\vs, \cdot)$ specifies dampening,
$\nabla\cdot\matr{H}_{\mathrm{E}}(\vs)\nabla u(\vs, \cdot)$ specifies anisotropic diffusion,
and $\nabla\cdot(\vomega_\mathrm{E}(\vs)u(\vs, \cdot)$ specifies advection.
This means that $\kappa_{\mathrm{E}}(\cdot)$ is a positive function,
$\matr{H}_{\mathrm{E}}(\cdot)$ is a differentiable spatially varying positive
definite $2\times 2$ matrix, and $\vomega_\mathrm{E}(\cdot)$ is a differentiable vector field.
The subscript E is short for ``Evolution''. We use zero flow boundary conditions given by
\begin{equation}
    \label{eq:diff_flow}
    \left(-\matr{H}_\mathrm{E}(\vs)\nabla u(\vs, t)+\vomega_\mathrm{E}(\vs)u(\vs, t)\right)\cdot\vn(\vs) = 0, \quad \vs \in \partial \mathcal{D}, \quad t\in(0,\infty),
\end{equation}
where $\vn(\cdot)$ is the outwards normal vector at the boundary.
To reduce the flexibility, we set $\kappa_\mathrm{F}(\cdot) = \kappa_{\mathrm{E}}(\cdot)$
and $\matr{H}_\mathrm{F}(\cdot) \equiv \matr{I}$ for the increments in $B(\cdot, \cdot)$.
This means that the coefficients to set or to estimate are $\kappa_\mathrm{E}(\cdot)$,
$\mathbf{H}_\mathrm{E}(\cdot)$, $\vomega_\mathrm{E}(\cdot)$, $\tau$, $\kappa_{I}(\cdot)$, and $\mathbf{H}_\mathrm{I}(\cdot)$. 




Other boundary conditions that could be considered for Equation \eqref{eq:advection_diffusion} are 
Dirichlet boundary conditions given by
\begin{equation}
    \label{eq:diff_dirichlet}
    u(\vs, t) = 0, \quad \vs \in \partial \mathcal{D}, \quad t \in (0, \infty).
\end{equation}
Additionally, for a rectangular domain, we could consider periodic boundary conditions where
the left-hand side and right-hand side are identified, and the bottom side and
top side are identified. In this context,
this means that what flows out on one side of the domain flows in on the other side.

\section{Modeling spatial-temporal data}
\label{sec:method}

\subsection{Latent model and observation model}
\label{subsec:obs_model}
We consider a bounded spatial domain $\mathcal{D}\subset\mathbb{R}^2$ and a bounded
temporal interval $\mathcal{T} \in \mathcal{T}$, and model the latent signal as
\[
    \eta(\vs, t) = \boldsymbol{x}(\vs, t)^\mathrm{T}\boldsymbol{\beta}+u(\vs, t),
    \quad \vs\in\mathcal{D}, \quad t\in\mathcal{T},
\]
where $\boldsymbol{x}(\cdot, \cdot)$ is a spatio-temporally varying $p$-dimensional
covariate vector, $\beta$ is a $p$-dimensional coefficient vector,
and $u(\cdot, \cdot)$ is the advection-diffusion-reaction model discussed in
Section \ref{subsec:advection_diffusion}. The parameters controlling $u(\cdot, \cdot)$
are discussed in Section \ref{subsec:parametrization}.

We assume $n$ measurements, $y_1,\dots, y_n\in\mathbb{R}$, are collected at spatial locations
$\vs_1, \ldots, \vs_n \in \mathcal{D}$ at times
$t_1, \ldots, t_n \in \mathcal{T}$. The observation model is
\begin{equation}
    y_i = \eta(\vs_i, t_i) + \epsilon_i, \quad i = 1, \ldots, n,
    \label{eq:obs_model}
\end{equation}
where $\epsilon_1, \ldots, \epsilon_n\overset{\text{iid}}{\sim}\mathcal{N}(0, \sigma_\mathrm{N}^2)$
with nugget variance $\sigma_\mathrm{N}^2 > 0$.

\subsection{Parametrizing the advection-diffusion-reaction model}
\label{subsec:parametrization}


The model described in Section \ref{subsec:advection_diffusion} has coefficients
$\kappa_\mathrm{E}(\cdot)$, $\mathbf{H}_\mathrm{E}(\cdot)$, $\vomega_\mathrm{E}(\cdot)$,
$\tau$, $\kappa_{I}(\cdot)$, and $\mathbf{H}_\mathrm{I}(\cdot)$.
Using the decomposition Equation in \eqref{eq:change_basis} on $\mathbf{H}_\mathrm{E}(\cdot)$
and $\mathbf{H}_\mathrm{I}(\cdot)$, we have one real number controlling the strength
of the forcing, $\log(\tau)$, and 10 real functions controlling
\begin{itemize}
    \item Evolution: $\log(\kappa_\mathrm{E}(\cdot))$, $\log(\gamma_\mathrm{E}(\cdot))$, $v_{\mathrm{E}, 1}(\cdot)$, $v_{\mathrm{E}, 2}(\cdot)$, $\omega_{\mathrm{E}, 1}(\cdot)$, and $\omega_{\mathrm{E}, 2}(\cdot)$
    \item Initial condition: $\log(\kappa_\mathrm{I}(\cdot))$, $\log(\gamma_\mathrm{I}(\cdot))$, $v_{\mathrm{I},1}(\cdot)$, and $v_{\mathrm{I}, 2}(\cdot)$.
\end{itemize}
The $\log$-transformation has been used for all quantities that must be positive.

\begin{figure}[!ht]
    \centering
    \includegraphics[width=0.7\textwidth]{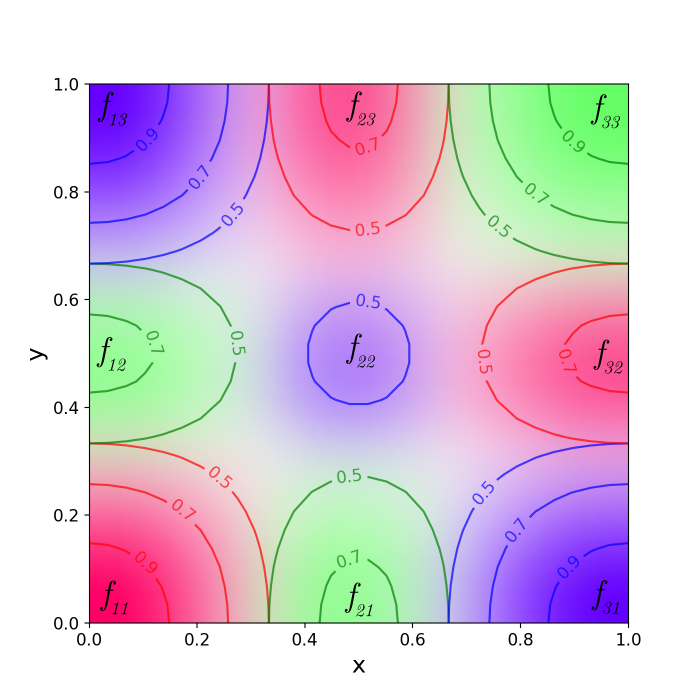}
    \caption{B-spline basis functions for a 2D domain. The colors indicate the
    value of each basis function, with different colors used solely to distinguish
    between the functions.}
    \label{fig:bspline}
\end{figure}

This means that if we have constant coefficients, there are in total 11 parameters
controlling the stochastic advection-diffusion-reaction. If we need spatially varying coefficients, 
we expand all 10 functions in spline function bases.
Specifically, in this paper, we use three 2nd-order
B-spline basis functions covering each spatial dimension. Then, combine using
tensor product splines as 
\begin{equation}
    \label{eq:tensor_spline}
f_{ij}(\vs) = B_{x,i}(x)B_{y,j}(y), \quad \mathrm{for} \enspace i = 1,2,3 \enspace \mathrm{and} \enspace j = 1,2,3,
\end{equation}
where $B_{x,i}(x)$ and $B_{y,j}(y)$ are the 2nd-order B-spline basis functions in the $x$- and
$y$-directions, respectively. A visualization of the B-spline basis functions is shown
in Figure~\ref{fig:bspline}. 
This is similar to the approach taken by \citet{fuglstad_does_2015} and
\citet{berild_spatially_2023} on the Whittle-Matérn SPDE. 
This gives in total 90 coefficients for the basis functions, and thus 91 parameters
including $\log(\tau)$. In general, one could use an arbitrary number of B-spline
basis functions in each dimension at the cost of increased computation time.
We refer readers to \citet[Appendix A.3]{berild_spatially_2023} for a detailed
description of the construction of 1D clamped B-splines.

In this paper, we use numerical simulations that are fully observed in space and time, and we have repeated realizations by considering numerical simulations from different time segments. For spatial non-stationarity, the work by \citet{berild_spatially_2023} indicates that dense observations and multiple realizations are necessary when estimating so flexible models without penalties on the parameters. As demonstrated in \citet{fuglstad_does_2015,fuglstad2019constructing}, one can add penalities or priors on the latent coefficients so that the model prefers the simpler stationary model and use sparsely observed data, but it is challenging to choose the penalization hyperparameters. Further work is needed for guidelines using flexible covariance structures in SPDE models.

We denote the parameter vector as $\boldsymbol{\theta}$, and the vector contains
$\log(\tau)$ and all coefficients of basis functions. When $\boldsymbol{\theta}$
is known, all functions are known, and the stochastic advection-diffusion-reaction
is fully specified.
We first discretize
the stochastic advection-diffusion-reaction model in space and time, then state the full hierarchical model and discuss computations. 


\subsection{Discretizing the advection-diffusion-reaction model}
\label{subsec:discretization}


We solve the advection-diffusion-reaction SPDE given in Equation~\eqref{eq:advection_diffusion}
numerically by first discretizing in time by backward Euler \citep{butcher2016numerical}, and then in space
using a finite volume method. This is inspired by \citet{clarotto_spde_2023}, but differ in that they used a FEM discretization in space.

We use a regular sequence of time points $0 = t_0^\mathrm{G} < t_1^\mathrm{G} < \ldots < t_{T-1}^\mathrm{G}$,
where the time step is $\Delta t$.
Backward Euler for Equation \eqref{eq:advection_diffusion}  gives
\begin{equation}
    \label{eq:adv_diff_tempDisc}
    (1+\Delta t\mathrm{A}(\cdot))u^{n+1}(\cdot) = u^n(\cdot) +\tau (B^{n+1}(\cdot)-B^n(\cdot)), \quad n = 0, \ldots, T-2,
\end{equation}
where $u^n(\cdot) = u(\cdot, t_n^\mathrm{G})$ and $B^n(\cdot) = B(\cdot, t_n^\mathrm{G})$
for $n = 0, \ldots, T-1$, and $u^0(\cdot)$ follows the initial condition
$\WM(\kappa_\mathrm{I}(\cdot), \matr{H}_\mathrm{I}(\cdot))$.
Note that
\[
    (B^1(\cdot)-B^0(\cdot))/\sqrt{\Delta t}, \ldots, (B^{T-1}(\cdot)-B^{T-1}(\cdot))/\sqrt{\Delta t} \overset{\text{iid}}{\sim}\WM(\kappa_E(\cdot), \matr{I})
\]
due to the restriction on the forcing made in Section \ref{subsec:advection_diffusion}.

The spatial domain $\calD$ is subdivided into $K$ regular rectangular grid
cells, each represented by a centroid $\vs_k$. We assume a piece-wise constant
solution with constant value within each grid cell, as is common in standard FVM.
A visualization of the spatial discretization is presented in Figure~\ref{app:fig:discretization}.
We refer to Appendix~\ref{app:sec:derivation} for technical details
and give an overall description of the results below. 

Let the discrete approximation of the solution at time $t_n$ be denoted by
\[
    \vu^n = \begin{bmatrix} u^n(\vs_1), \ldots, u^n(\vs_K)\end{bmatrix}^\mathrm{T}, \quad n = 0, \ldots, T-1.
\]
We can then discretize Equation \eqref{eq:adv_diff_tempDisc} in space as
\begin{equation}
    \label{eq:local_solution}
    \left(\matr{D}_\rV  + \Delta t\left(\mathbf{D}_\rV\matr{D}_{\kappa^2} -
     \matr{A}_\matr{H} +\matr{A}_{\bm{\omega}}\right)\right)\vu^{n+1} = 
      \matr{D}_\rV\vu^n + \tau\sqrt{\Delta t}\matr{D}_\rV\vphi^n,
\end{equation}
for $n = 0, \ldots, T-2$.
Here, $\vphi^0, \ldots, \vphi^{T-2} \overset{\text{iid}}{\sim} \mathcal{N}(\vzero,\matr{Q}_\mathrm{F}^{-1})$,
where $\matr{Q}_\mathrm{F}$ is a sparse matrix that arises from a discretization of
$\WM(\kappa_\mathrm{E}(\cdot), \matr{I}_2)$ using the method described in \citet{fuglstad_exploring_2014}.
Further, the initial state, $\vu^0$, is independent of $\vphi^0, \ldots, \vphi^{T-2}$,
and $\boldsymbol{u}^0\sim\mathcal{N}(\boldsymbol{0}, \matr{Q}_\mathrm{I}^{-1})$,
where $\matr{Q}_\mathrm{I}$ is a sparse matrix that arises from a discretization of
$\WM(\kappa_\mathrm{I}(\cdot), \matr{H}_\mathrm{I}(\cdot))$ using the method
described in \citet{fuglstad_exploring_2014}.
The matrices are
\begin{itemize}
    \item $\matr{D}_\mathrm{V} = V\matr{I}_K$: a diagonal matrix containing the areas, $V$, of the cells;
    \item $\matr{D}_{\kappa^2}$ and $\matr{A}_\mathrm{H}$ provide dampening and diffusion, respectively, and are discussed in \citet[Appendix~A.3]{fuglstad_exploring_2014};
    \item $\matr{A}_{\vomega}$: uses an up-wind scheme \citet[Section~20.3]{eymard2000finite} for advection and details are provided in Appendix \ref{app:sec:derivation}.
\end{itemize}

\subsection{Discrete hierarchical model}
\label{subsec:model_hierarchy}

Let $\vy = (y_1,\dots,y_n)^\mathrm{T}$ be the observations in Section \ref{subsec:obs_model}
at locations $\vs_1,\dots,\vs_n\in\mathcal{D}$ and times $t_1,\dots,t_n\in\mathcal{T}$.
Let
\[
    \vu = \begin{bmatrix} \vu^0 \\\vdots\\ \vu^{T-1}\end{bmatrix}
\]
be the vector of the discrete solution of the stochastic advection-diffusion-reaction
equation at all time grid points. We assume a piece-wise constant approximation so that for an arbitrary location $\boldsymbol{s}\in\mathcal{D}$ and time point $t\in\mathcal{T}$, the location is allocated to its corresponding grid cell in the spatial discretization, and the time point is allocated to its corresponding half open time interval $t\in[t_k^\mathrm{G}, t_{k+1}^\mathrm{G})$ for the appropriate $k$.
For ease of description, we assume that all observation time points match one of
the time grid points. I.e., $\{t_1, \ldots, t_n\}\subset\{t_0^\mathrm{G}, \ldots, t_{T-1}^\mathrm{G}\}$.
In the simulation study and the application, this will always be the case. 

The model can then be written in vector form,
\begin{equation}
    \label{eq:linmod}
    \vy = \matr{X}\vbeta + \matr{E}\vu + \vepsilon,
\end{equation}
where $\matr{X}$ is a design matrix of the covariates, where each row $i$ is $\vx(\vs_i,t_i)^\mathrm{T}$,
$\vbeta$ is the coefficients of the covariates, $\vepsilon \sim \calN(\vzero,\sigma^2_\mathrm{N}\matr{I})$,
and $\matr{E}$ is a $n\times KT$ matrix that selects the correct grid cell in the
correct time grid point for each observation.
Note that $\matr{E}$ has exactly one $1$ on each row and all other elements are $0$.
 

We introduce a Gaussian penalty
$\vbeta \sim \calN\left(\vzero,V_\beta\matr{I}\right)$, where $V_\beta > 0$ is a fixed number,
so that the full hierarchical model is 
\begin{equation*}
\begin{aligned}
    \textrm{Stage 1: }& \vy \mid \vbeta, \vu, \sigma_\mathrm{N}^2 \sim \calN \left(\matr{X}\vbeta + \matr{E}\vu, \sigma_\mathrm{N}^2 \matr{I} \right)\\ 
    \textrm{Stage 2: }& \vu \mid \vtheta \sim \calN(\vzero , \matr{Q}^{-1})  \enspace, \quad
    \vbeta \sim \calN\left(\vzero,V_\beta\matr{I}\right),
\end{aligned}
\end{equation*}
where $\matr{Q}$ is a $KT\times KT$ block tridiagonal matrix with sparse blocks of size $K \times K$ as described in Appendix \ref{app:sec:derivation}.
Since $\vw$ and $\beta$ are jointly Gaussian given $\vtheta$, we collect them together as $\vz = [\vu^\mathrm{T},\vbeta^\mathrm{T}]^\mathrm{T}$, and also let 
$\matr{S} = [\matr{E},\matr{X}]$ such that the hierarchical model becomes
\begin{equation*}
\begin{aligned}
    \textrm{Stage 1: }& \vy \mid \vz, \sigma_\mathrm{N}^2 \sim \calN \left(\matr{S}\vz, \sigma_\mathrm{N}^2 \matr{I} \right)\\ 
    \textrm{Stage 2: }& \vz \mid \vtheta \sim \calN\left(\begin{bmatrix} \vzero \\ \vzero\end{bmatrix} , \begin{bmatrix} \matr{Q}^{-1} & 0 \\ 0 & V_\beta\matr{I}\end{bmatrix}\right) 
\end{aligned}
\end{equation*}

If the parameters $\vtheta$ and $\sigma_\mathrm{N}^2$ are known, and
\[
\matr{Q}_z = \begin{bmatrix} \matr{Q} & 0 \\ 0 & \matr{I}/V_\beta\end{bmatrix},
\]
then the conditional distribution of the latent field is
 \begin{equation}
    \label{eq:cond_dist}
    \vz | \vy, \vtheta, \sigma_\mathrm{N}^2 \sim \mathcal{N}( \vmu_\rC, \matr{Q}_\rC^{-1}),
\end{equation}
with the conditional precision matrix 
\begin{equation}
    \label{eq:cond_prec}
    \matr{Q}_\rC = \matr{Q}_z + \sigma^{-2}_\mathrm{N}\matr{S}^\rT\matr{S},
\end{equation}
and the conditional expectation
\begin{equation}
    \label{eq:cond_exp}
    \vmu_\rC = \sigma^{-2}_\mathrm{N}\matr{Q}_\rC^{-1}\matr{S}^\rT\vy.
\end{equation}
Using this conditional distribution, the point prediction of the true value $y^*$ at the unobserved location $\vs^*\in \mathcal{D}$ is given by
\begin{equation*}
    \mathbb{E}\left[\vy^* | \vy, \sigma_\mathrm{N}^2, \vtheta \right]= \matr{S}_\mathrm{P}\vmu_\rC,
\end{equation*}
where $\matr{S}_\mathrm{P}$ is the corresponding design matrix for the prediction location.
The predictive distribution $\vy^* | \vy, \sigma_\mathrm{N}^2, \vtheta$ is Gaussian, and
the variance can be found either through algorithms that compute a partial inverse
of $\matr{Q}_\mathrm{C}$ at a similar computational complexity as a Cholesky factorization \citep{col27}, estimated through simulation, or through matrix free algorithms such as in \citet{clarotto_spde_2023}.
When forecasts are desired, one can either include additional unobserved time steps in the discretization, or simulate as above until the last time point with data and then simulate forward in time using the forward dynamics in Equation \eqref{eq:local_solution}.

\subsection{Parameter inference}
\label{subsec:param_inference}

The parameters of the models are inferred using maximum likelihood estimation
(MLE) or maximum a posteriori probability (MAP) estimation.
The log-posterior (or log-likelihood) is given by
\begin{equation}
\label{eq:posterior}
\begin{aligned}
\log \pi(\vtheta,\sigma^2_\mathrm{N}|\vy) &=  \mathrm{Const} + \log \pi(\vtheta,\sigma^2_\mathrm{N}) +
\frac{1}{2}\log |\matr{Q}_z| - \frac{1}{2}\log |\matr{Q}_\rC| + \\ &\phantom{=}- \frac{n}{2}\log \sigma^2_\mathrm{N} - 
\frac{1}{2}\vmu_\rC^\mathrm{T}\matr{Q}_{z}\vmu_\rC - \frac{1}{2\sigma^2_\mathrm{N}}(\vy-\matr{S}\vmu_\rC)^\rT(\vy-\matr{S}\vmu_\rC),
\end{aligned}
\end{equation}
where we use $\pi(\boldsymbol{\theta}, \sigma_\mathrm{N}^2) \propto 1$ in this paper
since we are emulating densely observed numerical model output and do not need extra regularization.
For the derivation of the log-likelihood, we refer the reader to \citet[Appendix A.4]{berild_spatially_2023}.


Inference is done using an optimization algorithm on the log-posterior to find 
the parameters that maximize it.
To aid the optimization, the gradient of the
log-posterior is used. 
Calculating the analytic gradient of the log-posterior requires computing 
$\Tr((\matr{Q}_z^{-1}-\matr{Q}_\rC^{-1})\frac{\partial \matr{Q}_z}{\partial \theta_i})$,
where $\Tr(\cdot)$ denotes the trace of the matrix. \citet{fuglstad_does_2015} took advantage of sparsity by using an algorithm that computes
a partial inverse only for the non-zero elements \citep{col27}, but computations are still expensive.
We use a stochastic trace estimator (STE), through the Hutchinson estimator \citep{hutchinson_stochastic_1990}, to
approximate the trace by
\begin{equation*}
    \widehat{\Tr(\matr{A})} = \widehat{\mathrm{E}\left[\boldsymbol{\xi}^\rT\matr{A}\boldsymbol{\xi}\right]} = \frac{1}{n}\sum_{i=1}^n \boldsymbol{\xi}_i^\rT\matr{A}\boldsymbol{\xi}_i,
\end{equation*}
where $n$ is an integer, and all elements of the vectors $\boldsymbol{\xi}_1,\ldots, \boldsymbol{\xi}_n$ are independent samples from the Rademacher distribution,
i.e., $\xi_{i,j} = \pm 1$ with equal probability for all $i$ and $j$. This approximation was also used in \citet{pereira_geostatistics_2022} and \citet{clarotto_spde_2023},
and is much faster than analytic calculations as it only requires $n$ matrix-vector products
and solves. 

The drawback of stochastic optimization is that the gradient is stochastic, and this can
lead to erratic behavior in the optimization, but it is often used because 
the computational speedup is so significant that the convergence is still faster.
There is a vast literature on stochastic gradient methods, and we refer the reader
to \citet{ruder_overview_2017} for an overview. 
Many of these methods tackle the problem of noisy gradients and use techniques
such as momentum, adaptive learning rates, and averaging of gradients to reduce
the noise. In this work, we use the Adam optimizer \citep{kingma_adam_2017} which
is a popular choice for optimization of neural networks and other large models.


\section{Example: reconstruction and forecasting}
\label{sec:examples}

In this section, we investigate the ability to reconstruct and forecast a complex
process for models of varying degrees of complexity.

\subsection{True process and observation model}

\begin{figure}[!ht]
    \centering
    \begin{subfigure}[b]{0.45\textwidth}
         \centering
         \includegraphics[width=\textwidth]{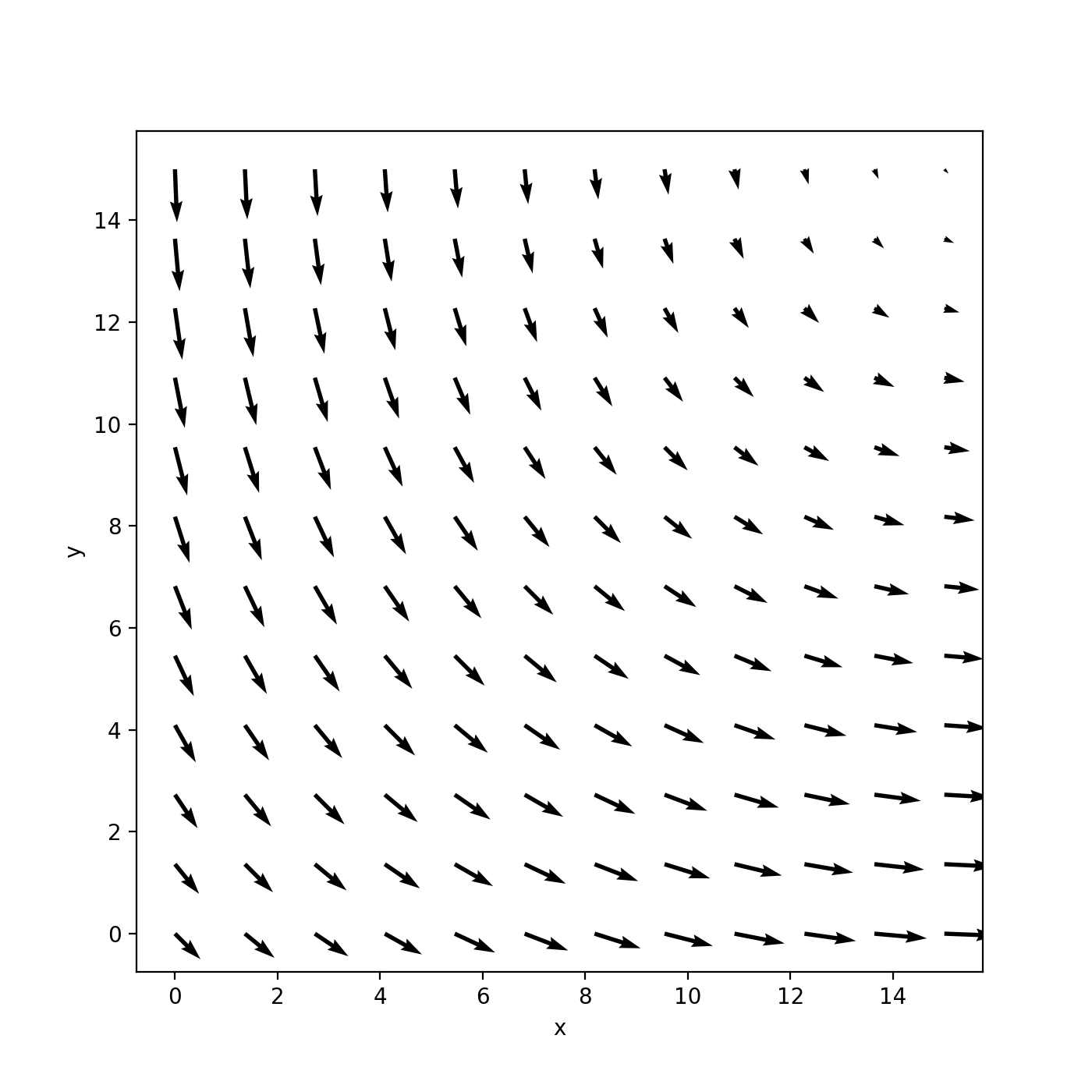}
         \caption{Advection\label{fig:sim_model1:advection}}
    \end{subfigure}
    \begin{subfigure}[b]{0.45\textwidth}
         \centering
         \includegraphics[width=\textwidth]{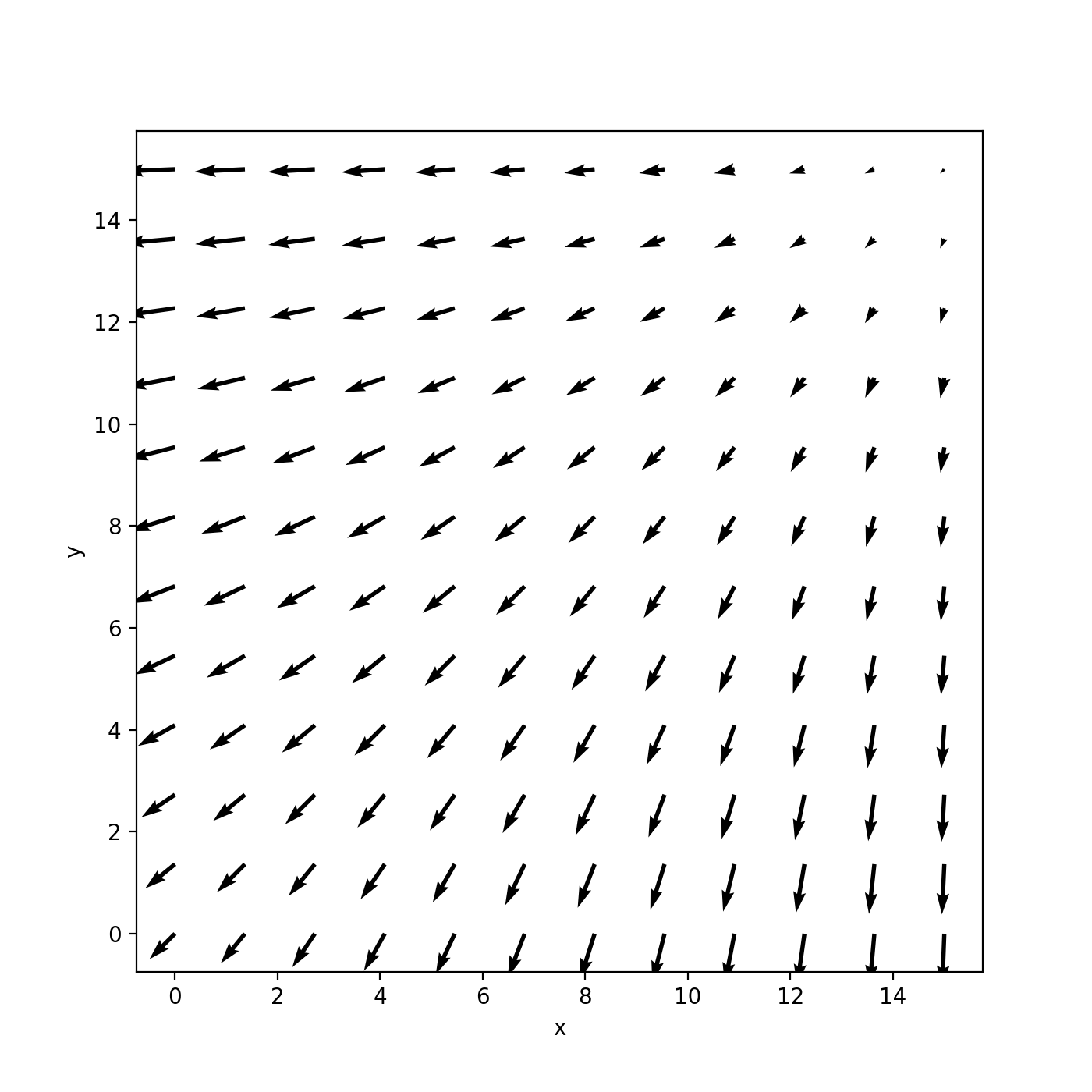}
         \caption{Extra diffusion}
    \end{subfigure}
    \caption{The unscaled vector fields describing (\textbf{a}) advection, and (\textbf{b})
    extra anisotropic diffusion. The length of the vectors shows the magnitude of the field.}
    \label{fig:sim_model1}
\end{figure}

Consider a spatial domain $\mathcal{D} = [0, 15]^2$ and temporal domain
$\mathcal{T} = [0, 2]$. We use a regular discretization with $M = 50$ grid cells
in $x$-direction, $N = 50$ grid cells in $y$-direction, and $T = 12$ timesteps. 
The true process has spatially varying advection and spatially varying anisotropic diffusion.
The advection vector field is
\begin{equation}
    \label{eq:sim_advection_par}
    \vomega(\vs) = 30 \begin{bmatrix}
        \sin\left((x/15 + 1/2)\pi/3\right)\cos\left((y/15 + 1/2)\pi/3\right) \\
        -\cos\left((x/15 + 1/2)\pi/3\right)\sin\left((y/15 + 1/2)\pi/3\right)
    \end{bmatrix},\quad \vs\in\mathcal{D},
\end{equation}
and the diffusion matrix is given by $\matr{H}(\cdot) = \exp(-1)\matr{I}+\vv(\cdot)\vv(\cdot)^\mathrm{T}$, where
\begin{equation}
    \label{eq:sim_diffusion_par}
    \vv(\vs) = 0.7 \begin{bmatrix}
        -\cos\left((x/15 + 1/2)\pi/3\right)\sin\left((y/15 + 1/2)\pi/3\right) \\
        -\sin\left((y/15 + 1/2)\pi/3\right)\cos\left((x/15 + 1/2)\pi/3\right)
    \end{bmatrix}, \quad \vs\in\mathcal{D}.
\end{equation}
Here, the vector fields are scaled with a factor of $30$ and $0.7$, respectively, which
gives a true process dominated by advection. 
A visualization of the unscaled vector fields is 
given in Figure \ref{fig:sim_model1}, where the length of the vectors shows the
magnitude of the fields.
For simplicity, the dampening coefficient is constant $\kappa(\cdot) \equiv \exp(-2)$,
the isotropic diffusion parameter is $\gamma(\cdot) \equiv \exp(-1)$,
and the scaling parameter for the forcing is $\tau = \exp(-4)$.
This defines the true spatio-temporal process
$\{u_{\mathrm{true}}(\vs, t): \vs\in\mathcal{D}, t\in\mathcal{T}\}$,
which we call NStat-True.

At the boundaries, we impose zero flow going in and out of the domain. As a consequence, diffusion inflates variance close to the boundaries, and variance is inflated near the boundary if there is strong advection into the boundary and variance is deflated if there is strong advection away from the boundary. To make behaviour more realistic within $\mathcal{D}$ and avoid large increases in variance, we add a buffer of $5$ grid cells to each side of the domain where we taper advection linearly to zero in the buffer. The new domain
$\mathcal{D}_\mathrm{B}$ then has $M=60$ and $N=60$ grid cells.

\begin{figure}[!ht]
    \centering
    \begin{subfigure}[b]{0.3\textwidth}
        \centering
         \includegraphics[width=\textwidth]{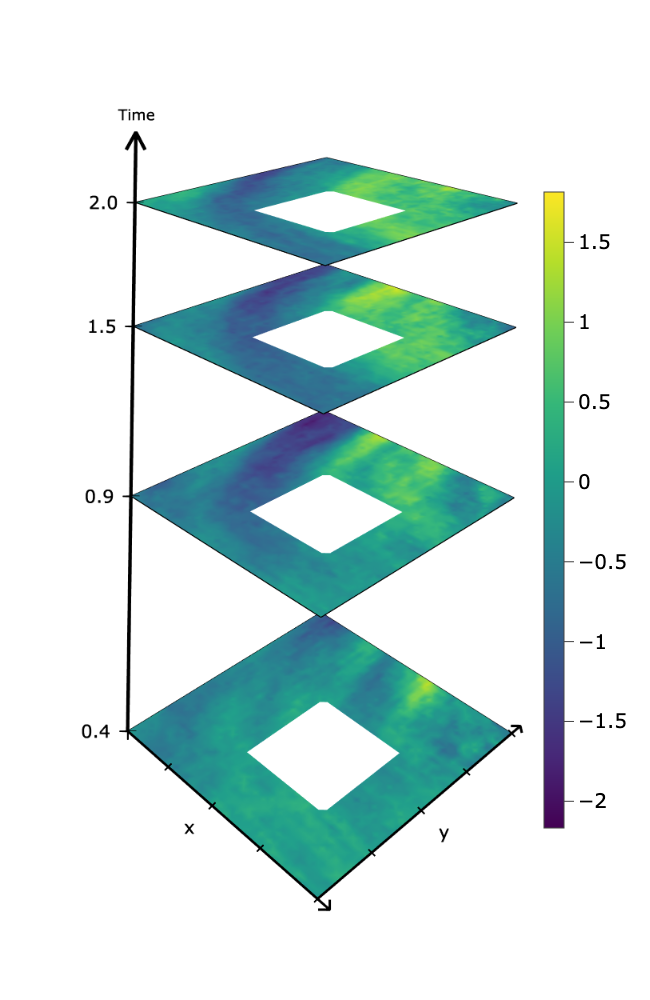}
         \caption{Realization}
    \end{subfigure}
    \begin{subfigure}[b]{0.3\textwidth}
        \centering
         \includegraphics[width=\textwidth]{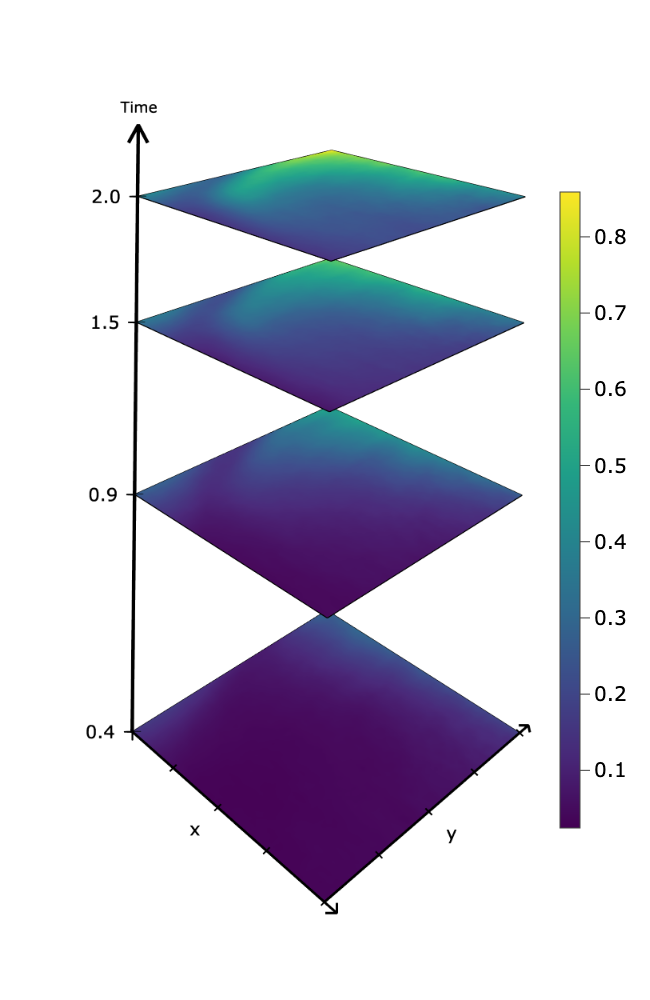}
         \caption{Variance}
        \end{subfigure}
    \begin{subfigure}[b]{0.3\textwidth}
         \centering
         \includegraphics[width=\textwidth]{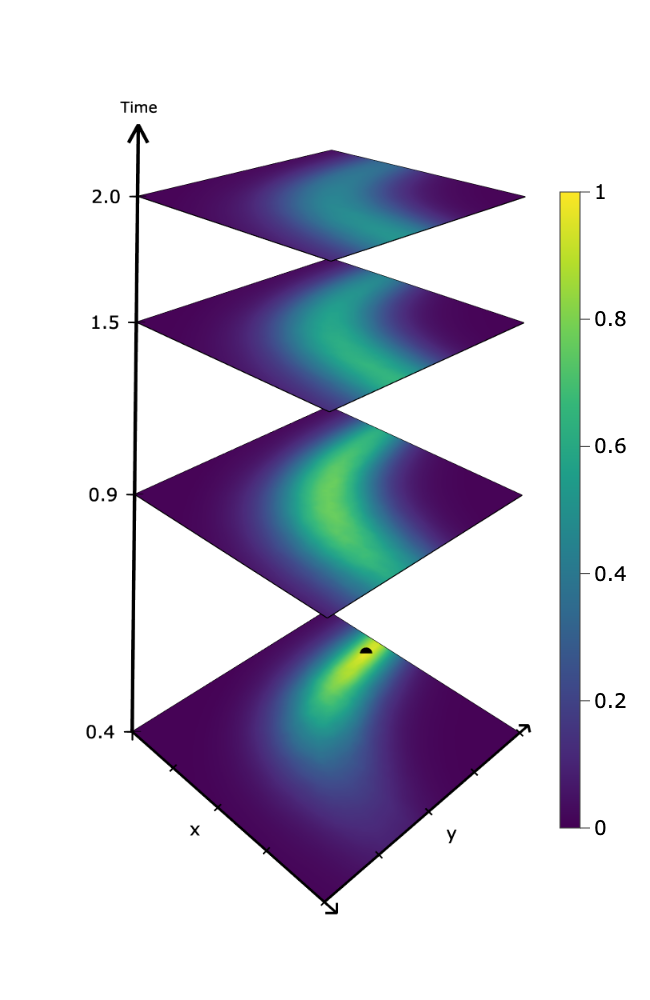}
         \caption{Correlation}
    \end{subfigure}
    \caption{Visualization of the model used to simulate 
    the data.}
    \label{fig:sim_model2}
\end{figure}

The field can be viewed as the spatio-temporal effect of the
concentration of a pollutant in a fluid, where the pollutant is advected by the
fluid with a favorable diffusion direction perpendicular to the advection direction.
The observation model in this study consists of a spatio-temporal effect observed with random noise.
Thus, Equation~\eqref{eq:obs_model} is simplified to
\begin{equation}
    \label{eq:sim_obs_model}
    y_i = u_{\mathrm{true}}(\vs_i, t_i) + \epsilon_i, \quad \epsilon_i \sim \mathcal{N}(0, \sigma_{\mathrm{N}}^2),
\end{equation}
for all cells $i$.
Here, the noise is i.i.d. Gaussian with variance $\sigma_{\mathrm{N}}^2 = 0.001$.
Figure~\ref{fig:sim_model2} shows one realization, marginal variance, and spatio-temporal
correlation with the marked location.

The missing rectangle of data is discussed in the next subsection.
Here, we observe the correlation move with the advection through time, and with a slight
diffusive effect. We can also see an advective effect in the realization, where 
higher or lower values areas are moved through space with time. The marginal variance
shows that the field is not stationary and it is highest at the entry point of the
advection vector field, and also the variance changes through time because the
initial field does not have the same variance as the field at the end.

\subsection{Goals, and training and test data}
\label{sec:subsec:example:goal}
We aim to mimic the emulation of a numerical model that has been densely observed with
a high signal-to-noise ratio. We generate 20 spatio-temporal realizations of NStat-True that
is used for training and testing. In each of the 20 realizations, the training data excludes the rectangular area $[6.1, 11.6] \times [3.7, 9.5]\subset \mathcal{D}$ (rounded to the first decimal), and the test data contains only the rectangular area. One realization of the training data is demonstrated in  Figure~\ref{fig:sim_model2}\textbf{a}. The goal is to assess the ability to reconstruct the missing rectangle, and to forecast for different lags forward in time.

There are $n_{\mathrm{train}}$ spatial training locations and $n_{\mathrm{test}}$ spatial test locations, and we can denote the training data and test data as $\boldsymbol{y}^{\mathrm{train}}_{t,r} = (y^{\mathrm{train}}_{1,t,r}, \ldots, y^{\mathrm{train}}_{n_{\mathrm{train}},t,r})^\mathrm{T}$ and $(y^{\mathrm{test}}_{1,t,r}, \ldots, y^{\mathrm{test}}_{n_{\mathrm{test}},t,r})^\mathrm{T}$, respectively, for time points $t=1, \ldots, 12$ and 
realizations $r = 1, \ldots, 20$. Parameters $\hat{\boldsymbol{\theta}}$ for each candidate model are estimated using all
training data $\boldsymbol{y}^{\mathrm{train}}_{t,r}$, $t = 1, \ldots 12$ and $r = 1, \ldots, 20$. We explain for a generic model and drop labels to distinguish between models for ease of presentation.

For each realization $r = 1, \ldots, 20$ for each
time point $t = 1, \ldots, 12$, we consider the $k$-step ahead point predictions
\[
    \hat{\boldsymbol{y}}^{\mathrm{test}}_{k,t,r} = \mathrm{E}[\boldsymbol{y}^{\mathrm{test}}_{t+k,r}|\boldsymbol{y}^{\mathrm{train}}_{t,r}, \hat{\boldsymbol{\theta}}], \quad k = 0, \ldots, 12-t.
\]
Note that $k = 0$ is a reconstruction of the missing rectangle for a given time point. The 
point predictions are evaluated using root mean square error (RMSE)
\begin{equation*}
    \mathrm{RMSE}_{k, t, r} = \sqrt{\frac{1}{n_{\mathrm{test}}}||\boldsymbol{y}^{\mathrm{test}}_{t+k,r} - \hat{\boldsymbol{y}}^{\mathrm{test}}_{k,t,r}||^2}, \quad t = 1, \ldots, 12-k, \quad r = 1, \ldots 20,
\end{equation*}
for $k = 0, \ldots, 12$, where $||\cdot||$ is the standard Euclidean distance. This gives 240 RMSE values for $k = 0$ and 20 RMSE values for $k = 12$.

We also assess the predictive distributions using the continuous ranked probability score (CRPS, \citet{gneiting_strictly_2007}),
\begin{equation*}
    \mathrm{CRPS}_{k,t,r} =  \frac{1}{n_{\mathrm{test}}}\sum_{i=1}^{n_{\mathrm{test}}} \sigma_{i,k,t,r}(z_{i,k,t,r} (2 \Phi(z_{i,k,t,r})- 1 ) + 2 \phi(z_{i,k,t,r}) - \pi^{-1/2}),
\end{equation*}
where $\sigma_{i,k,t,r}$ is the prediction 
variance for test location $i$ based on the predictive distribution $\boldsymbol{y}^{\mathrm{test}}_{t+k,r}|\boldsymbol{y}^{\mathrm{train}}_{t,r}, \hat{\boldsymbol{\theta}}$, and  $z_{i,k,t,r}$ is the difference between element $i$ in $\boldsymbol{y}^{\mathrm{test}}_{t+k,r}$ and element $i$ of the point forecast  $\hat{\boldsymbol{y}}^{\mathrm{test}}_{k,t,r} = \mathrm{E}[\boldsymbol{y}^{\mathrm{test}}_{t+k,r}|\boldsymbol{y}^{\mathrm{train}}_{t,r}, \hat{\boldsymbol{\theta}}]$, divided by $\sigma_{i,k,t,r}$. Here
$\Phi(\cdot)$ is the cumulative distribution function of the standard normal distribution,
and $\phi(\cdot)$ is the density function of the standard normal distribution.

For each $k = 0, \ldots, 12$,
we summarize RMSE and CRPS across all $t = 1, \ldots, 12-k$ and $r = 1, \ldots, 20$ using the average and the empirical standard deviation.

\subsection{Candidate models}
\label{subsec:candidate_models}
We consider four models:
\begin{itemize}
    \item NStat-True: The true covariance structure;
    \item NStat-AD: Stochastic advection-diffusion-reaction model discussed in
    Section \ref{sec:method} with 91 parameters to control covariance and 1 parameter for the nugget variance;
    \item NStat-Sep: Separable model in Section \ref{subsec:separable} with 37 parameters that control the covariance
    and 1 parameter for the nugget variance;
    \item Stat-AD: Stochastic advection-diffusion-reaction model discussed in
    Section \ref{sec:method} with spatially constant coefficients
    (11 parameters to control covariance) and 1 parameter for the nugget variance.
\end{itemize}
NStat-True gives the benchmark for how well we can do, NStat-AD is a flexible model
that we believe can capture the non-separability and non-stationarity well,
NStat-Sep is a separable model which includes non-stationarity,
and Stat-AD is a simplified non-separable model with spatially constant behaviour.

In all models, the mean structure is assumed to be zero with observation model
similar to Equation~ \eqref{eq:sim_obs_model}, but replacing $u_{\mathrm{true}}$
with the spatio-temporal effect of each respective model.


\subsection{Results}
\begin{figure}
    \centering
    \includegraphics[width=0.95\textwidth]{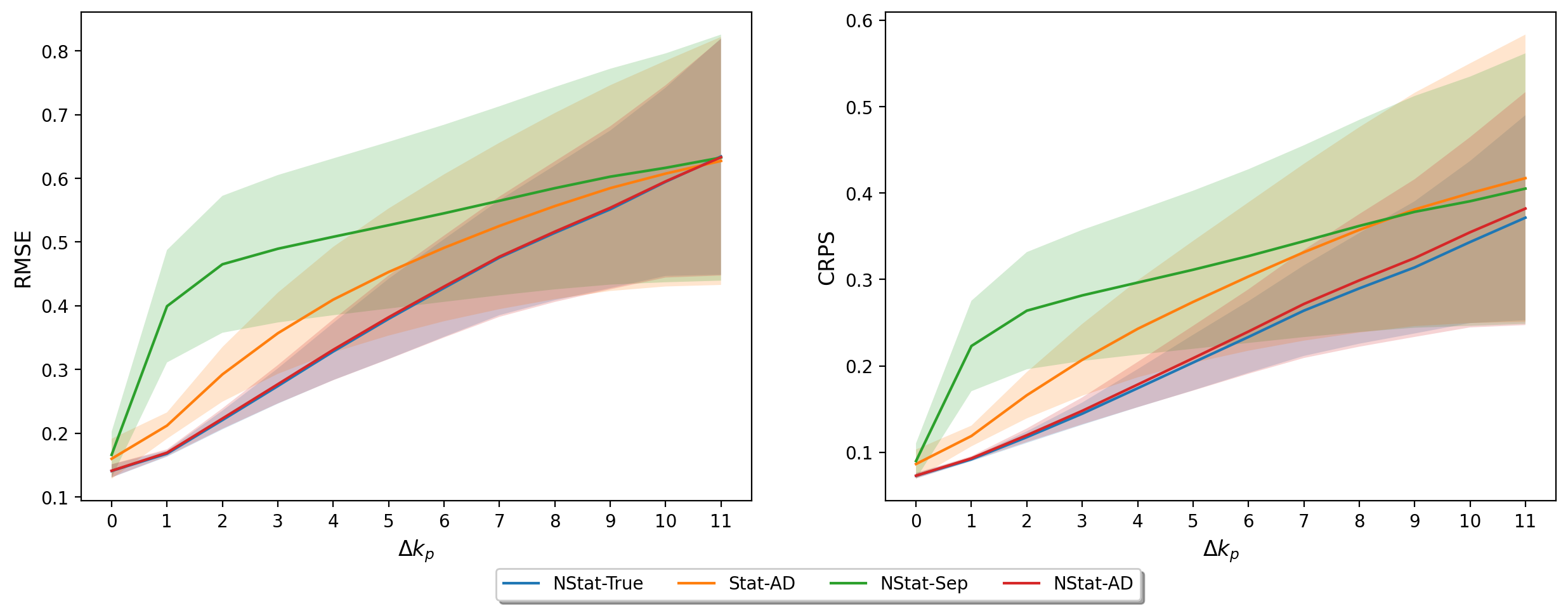}
    \caption{RMSE (left) and CRPS (right) for the reconstruction ($\Delta k_p=0$) and forecasting ($\Delta k_p>0$) of the
    different models. The solid line shows the average and shaded region shows $\pm$ one standard deviation. See Section \ref{sec:subsec:example:goal} for details.}
    \label{fig:simstud_adv_diff} 
\end{figure}

The results are shown in Figure~\ref{fig:simstud_adv_diff}.
Here, the $x$-axis represents the lag, $\Delta k_\mathrm{p} = t_\mathrm{p} - t_\mathrm{o}$, between the time step when predictions are made, $t_\mathrm{p}$, and the time step when data is observed, $t_\mathrm{o}$.
Reconstructions, $\Delta k_\mathrm{p} = 0$, only computes RMSE and CRPS for the test area, while forecasts $\Delta k_\mathrm{p} > 0$ computes RMSE and CRPS for both test and training locations.
For each candidate model and for each $\Delta k_\mathrm{p}$,  we take an average and an empirical standard deviation for the scores corresponding to legal time step combinations and realizations as discussed in Section \ref{sec:subsec:example:goal}.

First, we remark that NStat-AD has a very close fit to NStat-True, and their error curves are almost identical.
Further, all models perform relatively well in reconstructing the masked area,
but NStat-AD shows slight improvements over the other models in both RMSE and CRPS.
However, we can observe that the standard deviations are larger for reconstruction 
than for forecasting one timestep ahead. This could be expected as the reconstruction error
only considers the masked area, while the forecasting error considers the entire field.
The major difference is seen in the forecasting accuracy, already in the first timestep.
NStat-Sep is quite good in reconstruction, but the forecasting accuracy is poor. 
This is reasonable as its spatial covariance structure is flexible, but it is
unable to capture the strong advection present in the data-generating model.
NStat-Ad is better at forecasting than NStat-Sep. This is expected as it can
approximate parts of the advection even though it cannot capture the spatially
varying advection.
Based on Figure \ref{fig:sim_model1:advection}, it is reasonable that a constant
advection vector field could partially approximate the advection vector field.

\section{Emulating a numerical ocean model}
\label{sec:applications}

In this section, we evaluate the proposed approach for simulated AUV paths using
numerical ocean model output.

\subsection{Motivation and goal}
In this case study, we consider output from the SINMOD ocean model \citep{slagstad_modeling_2005}
developed by SINTEF Ocean.
The SINMOD ocean model is a 3D numerical model assembled by primitive equations of
fluids driven by atmospheric forces, freshwater input, and tides, that is solved
using finite difference methods on a regular horizontal grid of cell sizes
20km x 20km which is nested in several steps down to a minimum of 32m x 32m. 
The temporal resolution is 10 minutes and the depth layers are set specific
to a problem but often with higher resolution close to the surface to better
capture the variability.
The model deals with many ocean quantities such as temperature and currents.

The objective is to estimate a statistical emulator that captures the spatio-temporal
variations in the salinity field of a specific area of the Nidelva River outlet in
Trondheim, Norway, by utilizing output from the SINMOD model for one specific day (May 27, 2021).
This emulator will then be used to predict and forecast salinity levels based on
sparse measurements collected by a simulated AUV moving through 
the SINMOD data from other days than the day used in the estimation of the emulator.
By applying the emulator to different days for estimation and prediction, we aim
to assess its ability to adapt to varying conditions and simulate the transition
from a controlled, simulated environment to less predictable real-world ocean conditions.
We choose to consider time segments of 90 minutes which is a typical 
duration of AUV missions, and we want to be able to forecast within this same time frame.

\begin{figure}[!ht]
    \centering
    \begin{subfigure}[b]{0.3\textwidth}
         \centering
         \includegraphics[width=\textwidth]{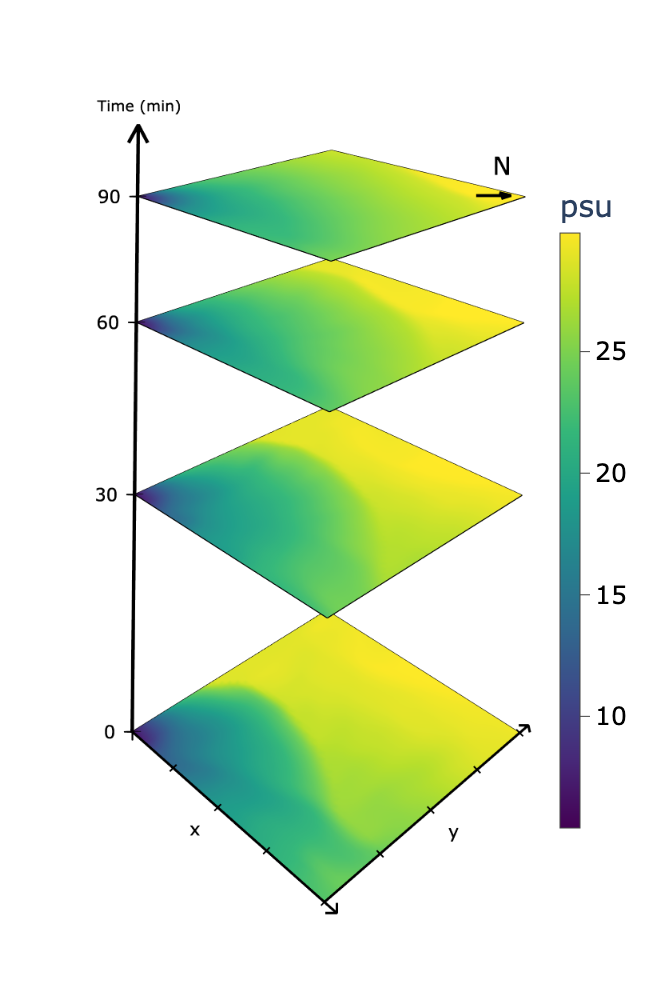}
         \caption{}
    \end{subfigure}
    \begin{subfigure}[b]{0.3\textwidth}
         \centering
         \includegraphics[width=\textwidth]{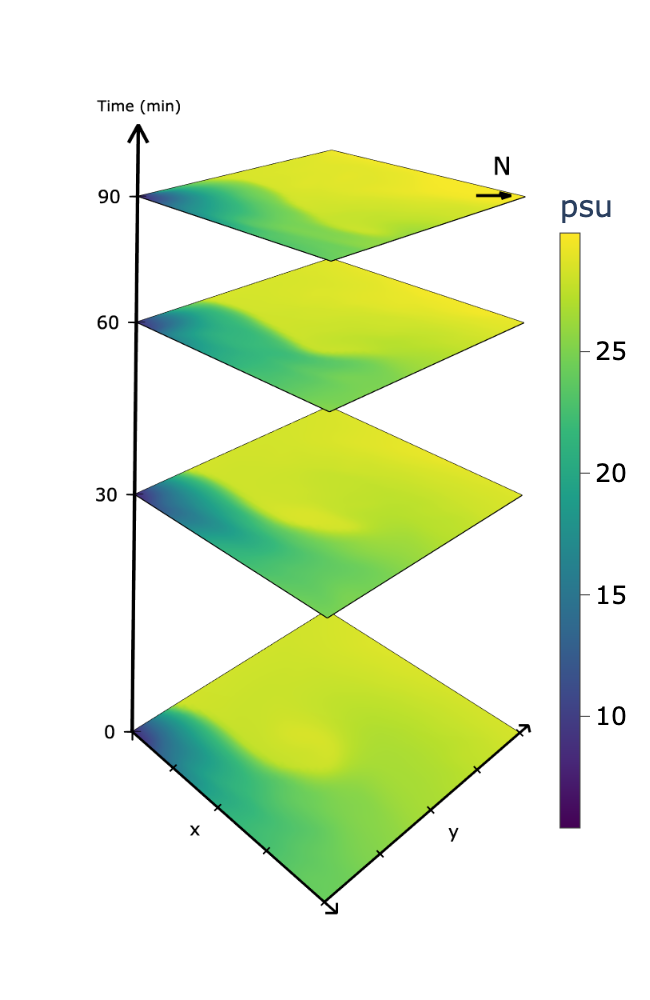}
         \caption{}
    \end{subfigure}
    \begin{subfigure}[b]{0.3\textwidth}
         \centering
         \includegraphics[width=\textwidth]{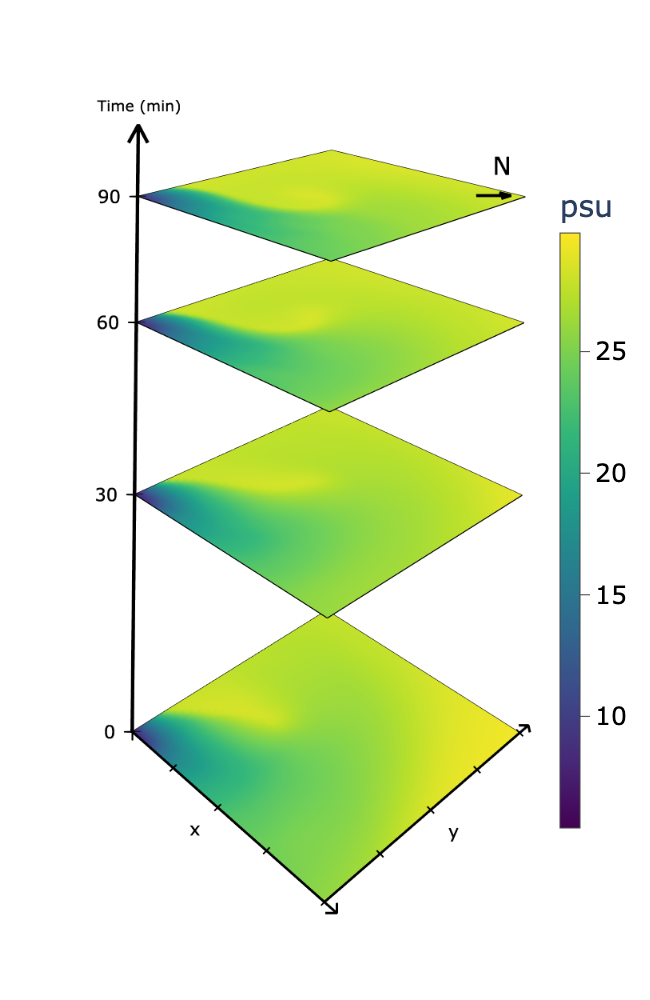}
         \caption{}
    \end{subfigure}
    \caption{Visualization of the salinity data generated by the numerical model (SINMOD).
    The three panels show three time segments of 90 minutes on May 27, 2021.
    Salinity is measured in PSU (practical salinity units) i.e., g/kg.}
    \label{fig:sinmod}
\end{figure}

Figure~\ref{fig:sinmod} shows three time segments of 90 minutes on May 27, 2021.
The Nidelva River outlet is located in the left corner in the figure or south direction 
in the area of interest. Here, we can clearly observe the freshwater input from the river
and spatio-temporal variations in the salinity field. 


The goal is to compare the proposed complex non-separable model to a separable model to
evaluate how much is gained by a non-separable model.
In this application, the true physical process (and the numerical model) clearly
has diffusion and advection, which gives a good platform
to evaluate the proposed approach.
Previous works have in general employed purely spatial GRFs onboard the AUVs
for real-time prediction, interpolation, and monitoring \citep{fossum2021learning,yaolin2023,berild_efficient_2024}.
However, a strong disadvantage of previous work is that the ocean is dynamic, and assuming that the
field is constant in time is not realistic for the typical duration of an AUV mission. Purely spatial models are usually used because they speed up onboard computations compared to spatio-temporal models.

We use paths that are pregenerated randomly, and treat the path as deterministic. The paths could also be chosen adaptively \citep{berild_efficient_2024}, but since our main focus is to evaluate prediction accuracy, we do not bring in further complications that can make it harder to interpret differences between models.



\subsection{Estimating the surrogate models}
\label{subsec:app_estimation}
The area of interest spans $1568\,\mathrm{m} \times 1408\,\mathrm{m}$, and we use
a discretization of $M = 50$ cells in $x$-direction and $N = 45$ cells in $y$-direction,
which gives a total of $K = M\cdot N = 2250$ spatial locations.
In addition, a buffer around the field is added to mitigate the effect of the boundaries on the marginal variances
of the field, and the total size is $M = 60$ cells in $x$-direction and $N = 55$ cells
in $y$-direction. This corresponds to a $160\,\mathrm{m}$ buffer. We also add advection tapering within this buffer 
to mitigate the effects of the boundaries even further.
This area is large enough to capture the dynamics of the river outlet and the
surrounding area, and with a number of timesteps that would be impractical to handle
with classical spatio-temporal models.

We use May 27, 2021 for estimating the model. We extract the data from SINMOD and add i.i.d. Gaussian noise with variance $\sigma_{\mathrm{obs}}^2 = 0.01$ to mimic that the truth is less smooth than numerical simulations from SINMOD. We call the resulting data  as the \emph{training data}.
The data consists of 144 timesteps, where each time step is 10 min,
and we split the dataset into $9$ non-overlapping time segments of $T = 9$ timesteps
(90 minutes) with one hour between each segment. The length of the time segments was chosen because we aim to estimate a surrogate model that works well for missions of length 1.5 hours. We do not use a moving window approach covering all windows of length 1.5 hours, and instead aim to cover a range of different behaviours without considering all possible 135 moving windows of length 9. Direct computations with 144 time steps would not be possible with Cholesky factorizations. In that case, it would be worthwhile to investigate matrix-free approaches such as discussed in \citet{clarotto_spde_2023}, but they did not consider problems as large as 144 time steps in the paper.

The training data is the true salinities 
$\vy_m = (\vy_{m,1}^\rT, \ldots, \vy_{m,T}^\rT)^\rT$, where  $\vy_{m,t} = (y_{m,t,1}, \ldots, y_{m,t,K})$ are the true salinities for time step $t$ in time segment $m$, for $t = 1, \ldots, T$ and $m = 1, \ldots, 9$.
To reduce effects of temporal evolution at a longer time scale than the envisioned mission lengths of $1.5$ hours, we subtract
the temporal mean of each time segment from the data as $\tilde{\vy}_m = \vy_m - \bar{\vy}_m$,
where $\bar{\vy}_m$ is the temporal mean for each location in space,
\begin{equation*}
     \bar{\vy}_m = \frac{1}{T}\sum_{t=1}^T \vy_{m,t}, \quad m = 1, \ldots, 9.
\end{equation*}

The surrogate models are then estimated on the resulting residuals, $\tilde{\vy}_1,\ldots,\tilde{\vy}_9$, 
using the model 
\begin{equation*}
     \tilde{\vy}_m = \matr{E}\vu_m + \vepsilon_m, \quad m = 1,\ldots, 9, 
\end{equation*}
where $\matr{E}$ links data in the $50\times 45$ area of interest with the $60\times 55$ latent spatial variables, $\vu_1, \ldots, \vu_9|\vtheta \overset{\text{iid}}{\sim} \calN(\vzero, \matr{Q}^{-1}(\vtheta))$, and $\vepsilon_1, \ldots, \vepsilon_9|\sigma_\mathrm{N}^2 \overset{\text{iid}}{\sim} \mathcal{N}(\vzero,\sigma_\mathrm{N}^2\matr{I})$. The model parameters
$\vtheta$  and $\sigma_\mathrm{N}^2$ are estimated to be $\hat{\vtheta}$ and $\hat{\sigma_\mathrm{N}^2}$, and  the estimated covariance structure of the surrogate model is $\matr{Q}(\hat{\vtheta})^{-1} + \hat{\sigma_\mathrm{N}^2}\matr{I}$.

We consider two of the models described in Section \ref{subsec:candidate_models}: 1) the proposed flexible stochastic diffusion-advection-reaction model
(NStat-AD), and 2) the complex non-stationary
separable model (NStat-Sep). As starting values in the optimization, we choose that advection should be similar to the currents,
which are generally pointed to the North (given by SINMOD output). The diffusion parameters 
are harder to choose, but are initially set perpendicular to the advection vector field.

\begin{figure}[!ht]
    \centering
    \begin{subfigure}[b]{0.45\textwidth}
         \centering
         \includegraphics[width=\textwidth]{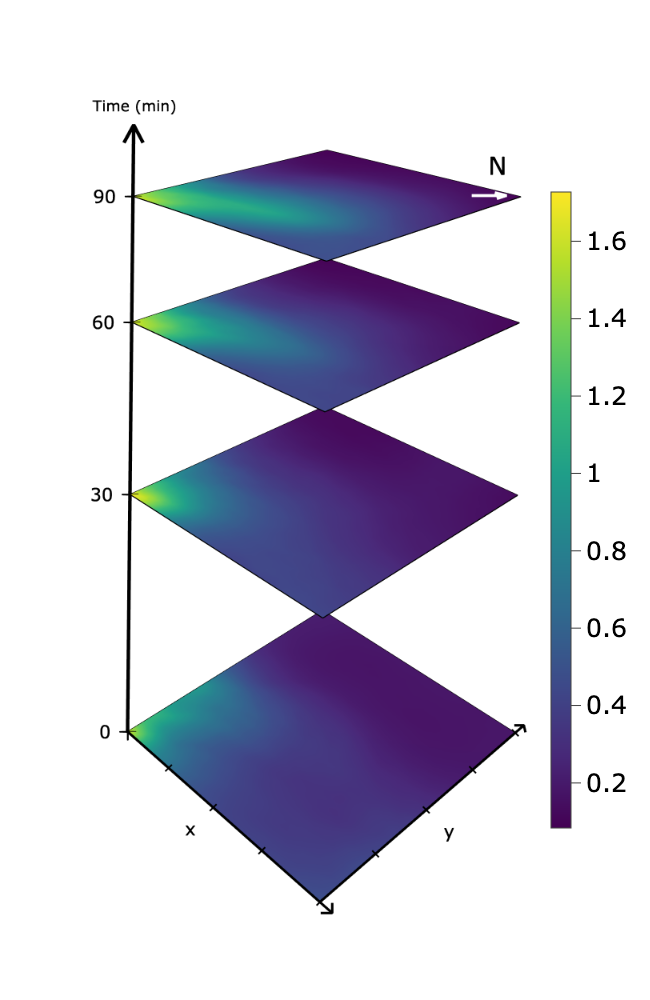}
         \caption{Variance}
    \end{subfigure}
    \begin{subfigure}[b]{0.45\textwidth}
         \centering
         \includegraphics[width=\textwidth]{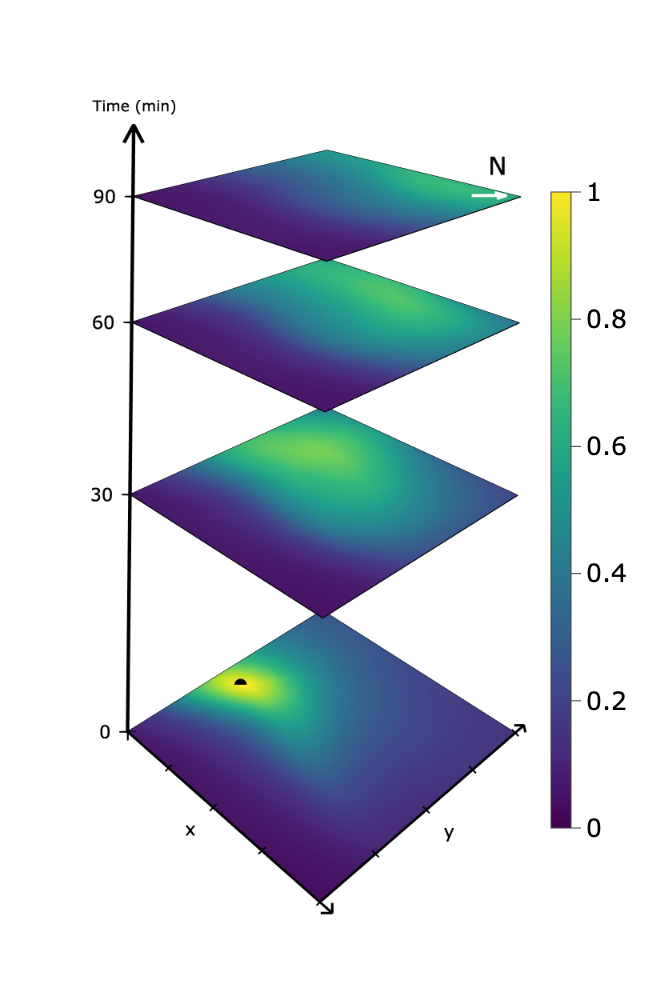}
         \caption{Correlation}
    \end{subfigure}
    \caption{The marginal variance (a) and the correlation (b) of the NStat-AD model
     estimated on SINMOD data from the 27th of May 2021.
     The marked location is the location in space and time where the correlation is visualized.}
    \label{fig:app_model}
\end{figure}

Figure~\ref{fig:app_model} shows the estimated marginal variance for NStat-AD,
and visualizes the estimated correlation with the marked location in space and time,
and all other locations.
Here, we can observe that the model has captured some spatially varying dynamics
in the SINMOD data. The correlation shows that there is an advection effect towards
the north and a possible diffusion effect towards the east (or west). We also observe
that the variance of the field is higher where the river outlet is located, which is
reasonable as the river will introduce varying freshwater that is affected by the
tides and the currents in the area.

\subsection{Evaluating the surrogate models}
The test dataset is comprised of numerical simulation data from SINMOD spanning nine
different days: May 4, May 10, May 11, May 28, and May 29 in 2021, and June 21, June 22,
September 8, and September 9 in 2022.
The setup for the test data mirrors that of the training data, and for each date, there are 9 time segments for a total of 72
time segments across the nine dates. The two estimated surrogate models in Section \ref{subsec:app_estimation}, NStat-Sep and NStat-AD, are used to model the test data. 

We consider a generic surrogate model and time segment, and write the estimated surrogate model for the truth as
\begin{equation}
    \boldsymbol{\eta} = \hat{\vmu} + \vu + \vepsilon,  
    \label{eq:modelSurrogate}
\end{equation}
where $\vu$ has the covariance matrix $\matr{Q}^{-1}(\hat{\vtheta})$ estimated in  Section \ref{subsec:app_estimation}, $\vepsilon\sim\mathcal{N}(\boldsymbol{0}, \hat{\sigma_\mathrm{N}^2}\matr{I})$ has the estimated variance $\hat{\sigma_\mathrm{N}^2}$ from Section \ref{subsec:app_estimation}, and  $\hat{\vmu} = (\hat{\vmu}_\rS^\rT, \ldots, \hat{\vmu}_\rS^\rT)^\rT$ is a known temporally constant mean with a spatially varying $\hat{\vmu}_\rS$ that is repeated $9$ times.
The spatially varying mean $\hat{\vmu}_\rS$ represents the expected spatial pattern, and is needed because the surrogate models in Section \ref{subsec:app_estimation} are estimated based on residuals from a spatially varying mean. 
We compute $\hat{\vmu}_\rS$ as the temporal average, across the 144 time steps in the training data in Section \ref{subsec:app_estimation}, for each spatial location. 
In real-world applications, it is hard to develop surrogate models
that are specifically tailored for the exact dates and duration of upcoming missions.
Therefore, we use this setup which evaluates applying the surrogate model to different days than used to estimate the surrogate model.
This serves as an effective way to assess the surrogate models' adaptability and
predictive accuracy.

\begin{figure}
     \centering
     \begin{subfigure}[b]{0.45\textwidth}
          \centering
          \includegraphics[width=\textwidth]{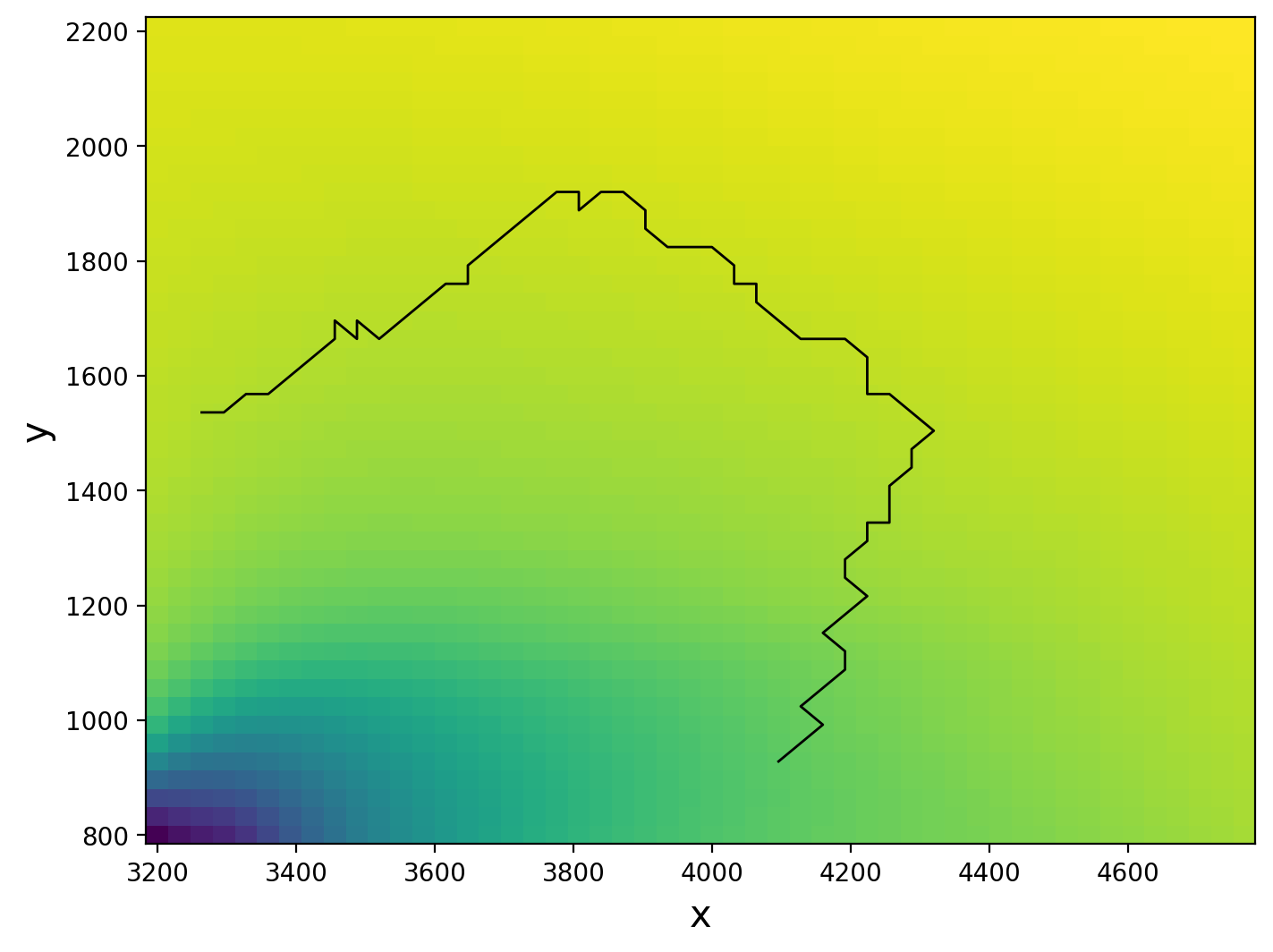}
          \caption{One path\label{fig:auv_paths:one}}
     \end{subfigure}
     \begin{subfigure}[b]{0.45\textwidth}
          \centering
          \includegraphics[width=\textwidth]{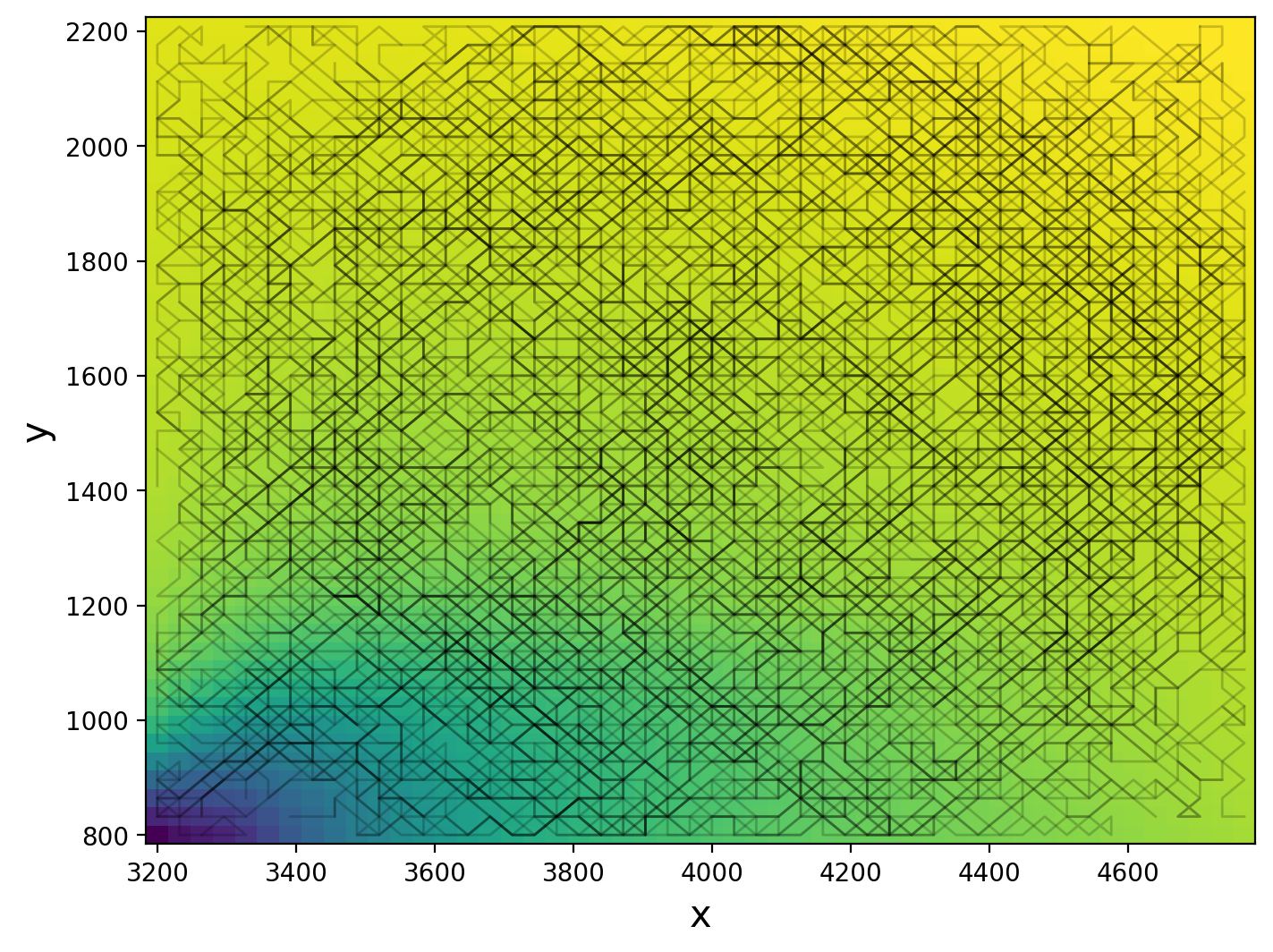}
          \caption{All paths\label{fig:auv_paths:all}}
     \end{subfigure}
     \caption{Illustation of one randomly generate AUV path (a) and all 200 paths (b)
     used in the evaluation of the surrogate models.}
     \label{fig:auv_paths}
\end{figure}

We consider an AUV that moves at a constant speed of $0.5\, \mathrm{m/s}$, and we create preplanned paths for the AUV by simulating 
random walks that minimize path intersections,
favor straight-line movement, and remain within the designated area of interest.
An example of such a generated path is depicted in Figure~\ref{fig:auv_paths:one}.
In practical scenarios, AUVs would collect data at specific intervals;
however, for this simulation, we assume data collection occurs whenever the AUV
enters a new grid cell or a new time step begins. 
This data collection procedure is similar to the result of data assimilation
done with FVM approach for the spatial SPDE model in \citet{berild_efficient_2024}. We generate 200 such paths, and use the same paths for each time segment. The paths are illustrated in Figure~\ref{fig:auv_paths:all}. This gives $72\cdot 200 = 14400$ combinations of time segments and paths. To simplify the description that follows, we term each such combination as a \emph{mission}.

For each of the 14400 missions, we assume that values are observed exactly. Typically a new measurement variance would need to be estimated from a trial run in the ocean based on empirically observed variation within grid cells \citep{berild_spatially_2023}. However, in this case the resolution matches for the surrogate model and the real-world variables.
The data collected in mission $i$ is denoted $\vz_i$ and is of length $n_{\mathrm{obs}, i}$, $i = 1, \ldots, 14400$.

All missions are evaluated separately, and we describe a generic mission and skip the index $i$ for ease of reading. We link the observation to the latent variables through $\vz = \matr{E}_{\mathrm{obs}}\boldsymbol{\eta}$,
where the matrix $\matr{E}_{\mathrm{obs}}$ links the $n_{\mathrm{obs}}$ observed locations in space and time to $\boldsymbol{\eta}$. The multivariate Gaussian surrogate model in Equation \eqref{eq:modelSurrogate} is then updated using the observed data. This gives a new multivariate Gaussian distribution $\boldsymbol{\eta}|\vz$, which describes posterior beliefs about the true states at all the 9 time steps.
The posterior mean, $\mathrm{E}[\boldsymbol{\eta}|\vz]$ serves as the point predictions for all locations in all time steps, and is assessed by computing the RMSE for the unobserved spatio-temporal locations during the mission. Similarily, we compute the an average CRPS across the unobserved spatio-temporal locations based 
on the marginal distributions from $\boldsymbol{\eta}|\vz$.

For each of the 72 time segments, we then compute the mean and the standard deviation of the RMSE values and CRPS values across the 200 missions in that time segment. This is done for each surrogate model.

\subsection{Results}
\label{subsec:app_results}

\begin{figure}[!ht]
     \centering
     \includegraphics[width = \textwidth]{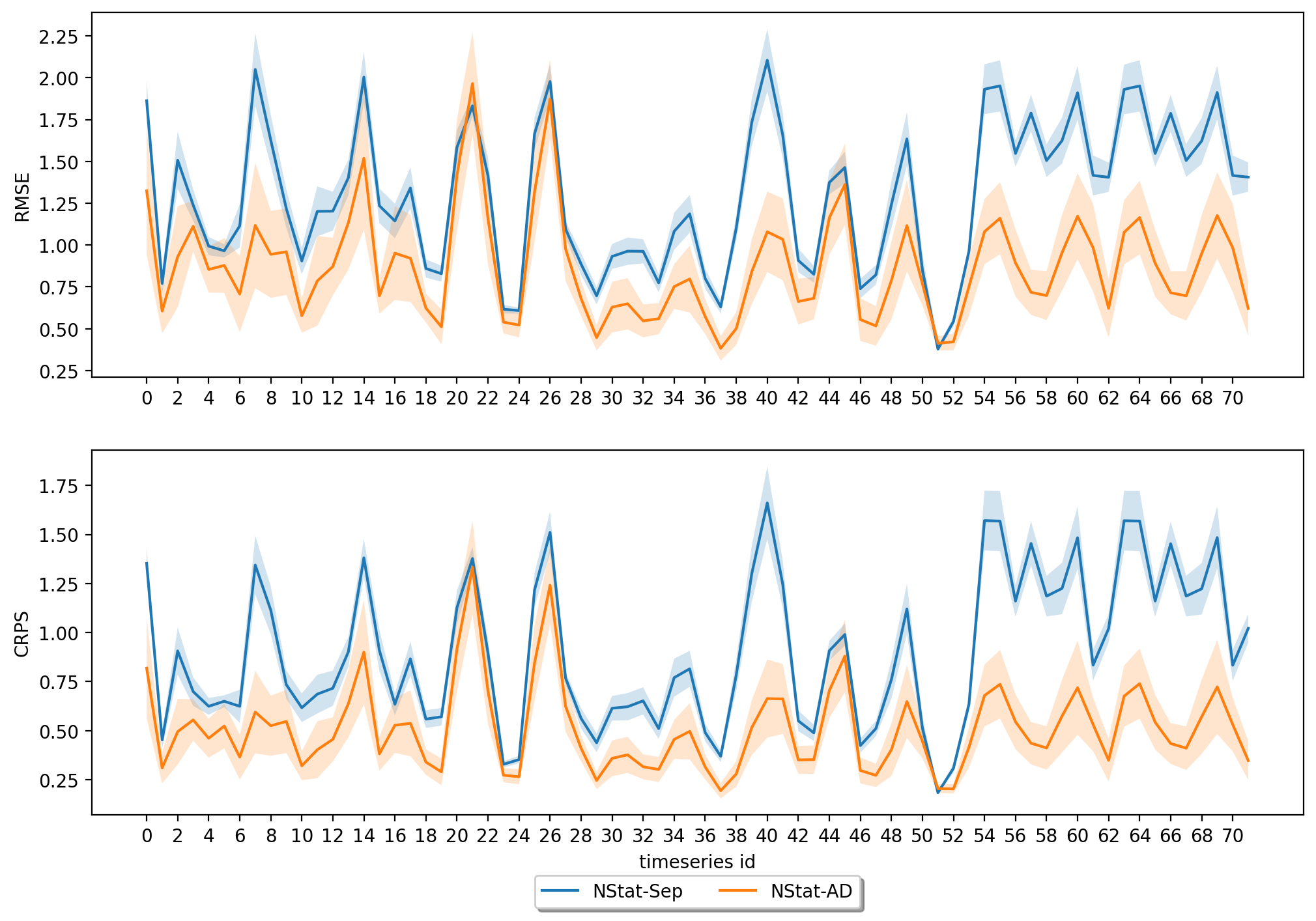}
     \caption{The RMSE (a) and the CRPS (b) for the 72 datasets used in the evaluation of the statistical emulators.
     The shaded regions are the standard deviations of the respective metrics across the 200 paths. 
     The x-axis shows the different datasets and the y-axis shows the RMSE and the CRPS.
     The blue line represents the NStat-AD model, and the orange line represents the NStat-Sep model.}
     \label{fig:app_eval}
\end{figure}
The comparative evaluation of the two surrogate models is depicted in Figure~\ref{fig:app_eval}.
The results indicate that the NStat-AD model generally outperforms the NStat-Sep
model in terms of both RMSE and CRPS across most time segments, with a few exceptions where
the NStat-Sep model performs marginally better.
This divergence is likely due to instances where the conditions significantly
deviate from those in the training data, causing the more complex NStat-AD model
to yield less accurate predictions.


The error bands, representing the standard deviations of the RMSE and CRPS across
the 200 paths, show greater variability for the NStat-AD model compared to
the NStat-Sep model.
This increased variation is anticipated given the higher accuracy of the NStat-AD
model, making it more sensitive to less informative AUV paths.
Despite this variability, the NStat-AD model consistently outperforms the
NStat-Sep model overall, even when considering this variation.

\subsection{Sensitivity to model choices}
\label{subsec:sensstud}
Several key choices were made when construction the model: 1) the spatial resolution, 2) the choice of flexibility in the covariance structure, and 3) the treatement of the boundary. Below we discuss the rationale behind the choices, and test sensitivity to some important choices.

The spatial resolution is chosen to match the spatial resolution ($32\,\mathrm{m}\times 32\,\mathrm{m}$) of the SINMOD data, and is the finest resolution resolvable from the data source. In this work we are not interested in evaluating methods for downscaling to higher resolution, and chose to use the same resolution for the surrogate model. If finer-scale models for the AUVs are desired, future work would be required to determine how to reliably downscale to finer resolutions.

\begin{figure}[!ht]
     \centering
     \includegraphics[width = \textwidth]{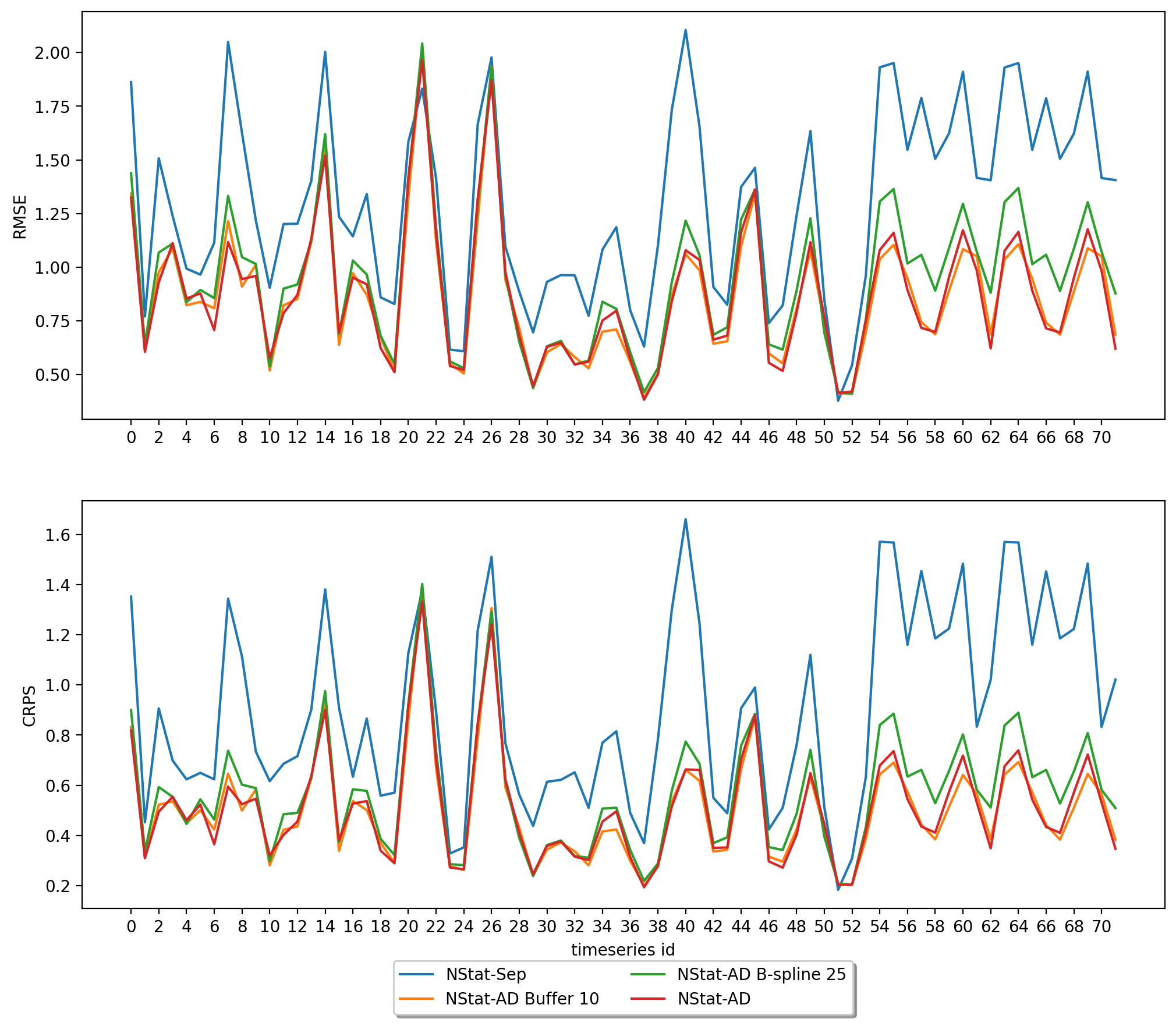}
     \caption{Sensitivity in Figure \ref{fig:app_eval} to changes to model choices.}
     \label{fig:sensstud}
\end{figure}

The flexibility with $3 \times 3$ basis functions for the coefficients in the SPDEs were chosen to give a moderate amount of flexibility that we expected could be estimated from the data. One might wonder if even higher flexibility would further improve the results compared to the stationary model. We repeated the steps also for a model with $5 \times 5$ basis functions (NStat-AD B-spline 25). Table \ref{tab:sensstudy} shows that the model still performs better than NStat-Sep, but that performance is worse than the less flexible NStat-AD. From Figure \ref{fig:sensstud}, it is clear that the more flexible model is consistently worse. This suggests that there may be a need for penalization of the flexibility to avoid overfitting to the training data. This is also  supported by the NStat-AD B-spline 25 model having larger maximum likelihood values than the NStat-AD on the training data. 

\begin{table}[!ht]
    \centering
    \caption{Scores under changes to model choices.    \label{tab:sensstudy}
}
    \begin{tabular}{l|c|c|c}
        Model & Extension & RMSE & CRPS ($10^{-1}$) \\
        \hline
        NStat-Sep & - & 1.376 & 9.05 \\
        NStat-AD & - & 0.947 & 5.06\\
        NStat-AD Buffer 10 & Larger buffer zone & 0.940 & 5.12\\
        NStat-AD B-spline 25& More flexibility in covariance & 1.026 & 5.69
    \end{tabular}
\end{table}

Alternative boundary choices such as fixed values on the boundary, and periodic boundary conditions are undesirable for the application. Fixed values on the boundary gives variances close to zero, and periodic boundary conditions allows advection to transport small/large values from one side of the domain to the other. Additionally, it is not straightforward to mitigate such effects in a non-stationary model since there is no natural extension of the model to the entire $\mathbb{R}^2$ without knowing the spatially varying coefficients also outside the domain $\mathcal{D}$. In our opinion, the no-flow boundary conditions coupled with the tapering of advection offers most reasonable approach to mitigate undesirable behaviour. To investigate the sensitivity to the choice of using a buffer of 5 cells ($160\,\mathrm{m}$), we repeated the steps for a model (NStat-AD Buffer 10) with 10 cells buffer ($320\, \mathrm{m})$). Table \ref{tab:sensstudy} shows a slight improvement in the third significant digit, but Figure \ref{fig:sensstud} shows that the improvement is not consistent across all time segments.

The results indicate some sensitivity to the choice of flexibility in the model, and less to the size of the buffer region.  Sensitivity to flexibility is not unexpected.  \citet{fuglstad_does_2015} has previously discussed that more flexibility is not always better, and that a key point is to put the extra flexibility in the correct place.

\section{Discussion}
\label{sec:discussion}
We are able to estimate spatially varying advection and diffusion by using bases
of tensor product B-splines for the coefficients in the SPDE.
The model with advection and diffusion performed better than a non-stationary
separable spatio-temporal model, which struggled with advection and diffusion.
In the simulation study, we even saw that the simplified non-separable model
outperformed the more complex non-stationary separable model. 

Stationary models are contained in the new model class by using one parameter per
coefficient instead of basis functions. The initial state of the spatio-temporal processes can be visualized through the marginal variances and correlations in the initial distribution for the GRF, and the dynamics can be understood through
the dampening, the diffusion tensor, and the advection vector field. This means that we can visualize the spatio-temporal dynamics by visualizing a spatially varying dampening, spatially varying ellipses showing the local strength of diffusion in each direction, and a vector field visualising the directions and strengths of advection. In this way, stochastic behaviour can be understood through physical phenomena such as diffusion and advection.
Further, in the advection-dominated simulation study, we do not observe the
instability seen by \citet{clarotto_spde_2023} for their FEM approximation.
This suggests that FVM is a promising alternative to FEM for this problem. 

In the simulation study, we found that while the complex models substantially outperformed
the simpler ones for forecasting, they only showed a marginal improvement in
reconstruction tasks. This suggests that correctly modeling the spatio-temporal dynamics
is more important for forecasting than for interpolation. 
The predictive performance of the complex model closely mirrored
that of the true model, affirming our ability to accurately estimate its parameters
even when a sub-area was missing.


The potential applications for this approach are vast, as advection and diffusion
are natural first approximations of spatio-temporal dynamics.
However, one may need more complex mechanisms to capture, for example, long-distance coupling. 
A key challenge for practical application lies in acquiring sufficient data to accurately estimate
the proposed model. In cases where data is more sparse, it will be necessary to
incorporate penalties or priors that limit the spatial variation of the coefficients in the SPDE.
Determining the appropriate strength of the penalties can be challenging for
non-stationary models \citep{fuglstad_does_2015}.
Before applying the proposed model to such settings, one would need to investigate
appropriate penalties and priors. 

An intriguing extension of our proposed approach involves the local refinement of
tensor product B-splines \citep{patrizi_adaptive_2020}.
This refinement adjusts the placement of knots within the B-splines depending on
where increased flexibility is required or where less is beneficial.
Another potential enhancement is the adoption of triangular instead of rectangular grid cells.
This modification would offer enhanced flexibility in the spatial domain,
allowing for more accurate representations of complex geographical features
like coastlines, islands, and fjords.
While these extensions would necessitate additional implementation efforts,
they present valuable opportunities for further research and development in future projects.

Optimizing our model presents several challenges and opportunities for improvement.
One notable limitation is that the vector field describing the anisotropic
diffusion exhibits two equivalent optima, complicating optimization efforts,
particularly since this vector field varies spatially.
Exploring alternative parameterizations of the vector field that circumvent this
issue such as \citet{llamazares2024parameterization} could be beneficial.
In our efforts to optimize the model parameters, we have utilized the ADAM optimizer \citep{kingma_adam_2017},
chosen for its effectiveness in handling vanishing gradients—a common issue in
machine learning scenarios where certain parameter gradients are extremely small.
We have tested several optimizers, including Rmsprop, Adagrad, and Adadelta \citep{ruder_overview_2017},
but none outperformed ADAM in our studies.
We also experimented with AdaHessian \citep{yao_adahessian_2021}, which scales gradients using the diagonal
of the Hessian matrix.
However, this approach did not enhance the optimization process, likely because
the parameters are too interdependent to benefit from scaling by only the diagonal elements.
Calculating the full Hessian matrix, while potentially more effective,
is prohibitively expensive in computational terms for this model.
Exploring these optimization challenges further could yield significant
advancements in the performance and applicability of our approach.

\section*{Author Contributions}
Martin Outzen Berild performed writing - original Draft, development of the software, and the methodology, and
Geir-Arne Fuglstad contributed with writing - original draft, conceptualization, and supervision.

\section*{Funding}
This work is funded by the Norwegian Research Council through the MASCOT project 305445.

\newpage
\appendix
\setcounter{section}{0}%
\renewcommand{\thesection}{\Alph{section}}
\setcounter{subsection}{0}%
\renewcommand{\thesubsection}{\Alph{section}.\arabic{subsection}}
\setcounter{table}{0}%
\renewcommand{\thetable}{S\arabic{table}}%
\setcounter{figure}{0}%
\renewcommand{\thefigure}{S\arabic{figure}}%
\setcounter{equation}{0}%
\renewcommand{\theequation}{S\arabic{equation}}%

\section{Discretizing the advection-diffusion SPDE}
\label{app:sec:derivation}

\subsection{Problem formulation}

In this part, we will consider the stochastic advection-diffusion-reaction equation,
\begin{equation}
    \label{app:eq:advection_diffusion}
        \frac{\partial }{\partial t}u(\bm{s},t) + (\kappa^2(\vs) -
        \nabla \cdot \matr{H}(\vs) \nabla)u(\bm{s},t) + \nabla \cdot (\vomega(\vs)u(\bm{s},t))
        = \tau \frac{\partial}{\partial t} B(\bm{s},t).
\end{equation}
for $\vs\in\mathcal{D}\subset\mathbb{R}^2$, where $\mathcal{D}$ is the spatial domain, and $t\in\mathcal{T}$, where $\mathcal{T}$ is the temporal domain.
Here, $\kappa(\cdot)$ is a positive function, $\matr{H}(\cdot)$ is a spatially varying $2\times2$ positive definite matrix, $\vomega(\cdot)$ is a vector field, $\tau > 0$ is a constant, and $B(\cdot, \cdot)$ is the Q-Wiener process defined in Section \ref{subsec:separable} through $\WM(\kappa_\mathrm{F}(\cdot), \matr{H}_\mathrm{F}(\cdot))$. Compared to Equation \eqref{eq:advection_diffusion}, we skip the subscript ``E'' to ease notation in the derivations that follow. We consider no flow boundary conditions,
\[
    \label{eq:app_diff_flow}
    \left(-\matr{H}(\vs)\nabla u(\vs, t)+\vomega(\vs)u(\vs, t)\right)\cdot\vn(\vs) = 0, \quad \vs \in \partial \mathcal{D}, \quad t\in\mathcal{T},
\]
where $\vn(\cdot)$ is the outwards normal vector at the boundary.

We assume:
\begin{itemize}
    \item the spatial domain is rectangular $\mathcal{D} = [A_1,A_2]\times[B_1,B_2]$, where $A_1 < A_2$ and $B_1 < B_2$;
    \item the temporal domain is $\mathcal{T} = (0, T_\mathrm{t}]$, where $T_\mathrm{t} > 0$ is the terminal time.
    \item the initial condition is $u(\cdot, 0) \sim \WM(\kappa_\mathrm{I}(\cdot), \matr{H}_\mathrm{I}(\cdot))$
\end{itemize}

We solve the advection-diffusion-reaction SPDE given in Equation~\eqref{app:eq:advection_diffusion}
numerically by first discretizing in time by backward Euler, and then in space
using a finite volume method. This is inspired by \citet{clarotto_spde_2023}, but differ in that they used a FEM discretization in space.

\subsection{Temporal discretization}
We use a regular temporal grid where $0 = t_0^\mathrm{G} < t_1^\mathrm{G}<\cdots<t_{T-1}^\mathrm{G} = T_\mathrm{t}$, and the time step length is $\Delta t$.
Let $u^n(\cdot) = u(\cdot, t_n^\mathrm{G})$ and $B^n(\cdot) = B(\cdot, t_n^\mathrm{G})$ for $n = 0, \ldots, T-1$. Then
using backward Euler on Equation \eqref{app:eq:advection_diffusion} gives
\begin{equation}
\label{app:eq:advection_diffusion_implicit_euler}
\begin{aligned}
 u^{n+1}(\cdot)  &= u^n(\cdot) \\
                 &\phantom{=}-\Delta t\left\{\left[\kappa(\cdot)^2 - \nabla \cdot \matr{H}(\cdot) \nabla\right]u^{n+1}(\cdot) 
+\nabla\cdot(\vomega(\cdot) u^{n+1}(\cdot))\right\} +\\
                 &\phantom{=}+ \tau(B^{n+1}(\cdot)-B^n(\cdot)),
\end{aligned}
\end{equation}
for $n = 0, 1, \ldots, T-2$. From the definition of $B(\cdot, \cdot)$, we have that
\[
    (B^1(\cdot)-B^0(\cdot))/\sqrt{\Delta t}, \ldots, (B^{T-1}(\cdot)-B^{T-1}(\cdot))/\sqrt{\Delta t} \overset{\text{iid}}{\sim}\WM(\kappa_\mathrm{F}(\cdot), \matr{H}_\mathrm{F}(\cdot)).
\]
The GRFs $u^0(\cdot), \ldots, u^{T-1}(\cdot)$ gives an approximation that is discrete in time, but continuous in space.

\subsection{Spatial discretization}
\begin{figure}[!ht]
    \centering
    \includegraphics[width=0.8\textwidth]{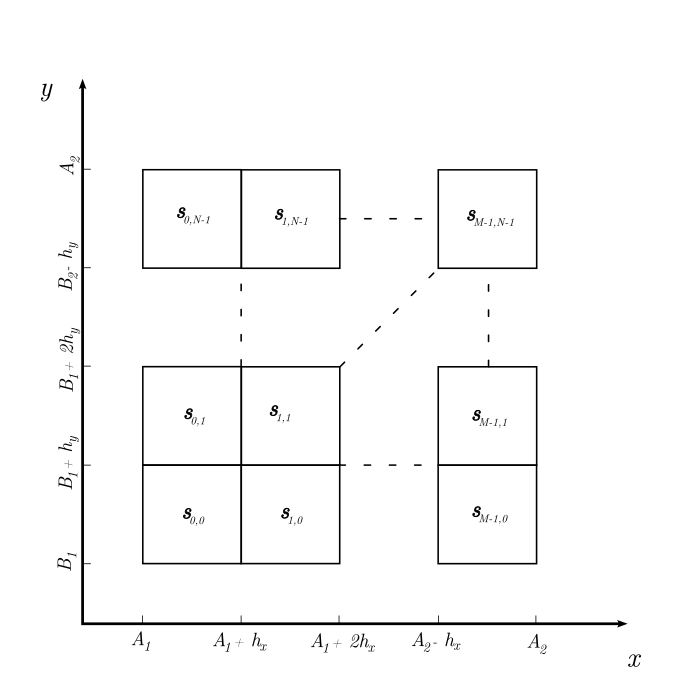}
    \caption{Visualization of the discretizated spatial domain.}
    \label{app:fig:discretization}
\end{figure}

Divide the rectangle $\mathcal{D} = [A_1,A_2]\times[B_1,B_2]$ into a regular grid
of $M$ cells in the $x$-direction and $N$ cells in the $y$-direction.
The grid cells are indexed by
$i = 0,\ldots,M-1$ and $j = 0,\ldots,N-1$ in the $x$- and $y$-direction, respectively.
The grid cell centers are denoted by $\vs_{i,j} = (x_i,y_j)$, and the side length of the grid cells are denoted by
$h_x = (A_2 - A_1)/M$ and $h_y = (B_2 - B_1)/N$. A visualization of the discretized spatial domain
is shown in Figure~\ref{app:fig:discretization}. 

\begin{figure}[!ht]
    \centering
    \includegraphics[width=0.8\textwidth]{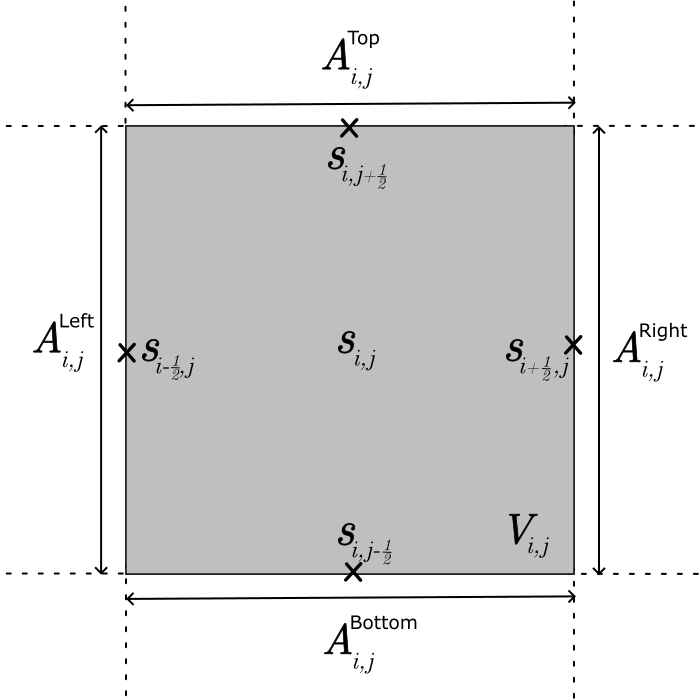}
    \caption{A visualization of a grid cell. The cell volume $V_{i,j}$ is shown in gray.}
    \label{app:fig:gridcell}
\end{figure}

The volume/area of each grid cell is denoted by $V_{i,j} = h_xh_y$. 
We denote the grid cell by $E_{i,j}\subset\mathcal{D}$ and the boundary
of the grid cell by $\partial E_{i,j}$. The boundary of each grid cell consists of four faces (or edges) denoted by $A_{i,j}^\rL$, $A_{i,j}^\rR$, $A_{i,j}^\rU$, and $A_{i,j}^\rD$. 
We denote the centers of the respective faces as $\vs_{i+1/2,j}$, $\vs_{i-1/2,j}$,
$\vs_{i,j+1/2}$, and $\vs_{i,j-1/2}$. See Figure~\ref{app:fig:gridcell} for a visualization.

We apply FVM and integrate the temporally discretized Equation \eqref{app:eq:advection_diffusion_implicit_euler} over a cell,
\begin{gather*}
        \int_{E_{i,j}} u^{n+1}(\vs)\mathrm{d}\vs = \int_{E_{i,j}} u^{n}(\vs)\mathrm{d}\vs  \\ 
        -\Delta t \left\{\int_{E_{i,j}}  \kappa^2(\vs)u^{n+1}(\vs)\mathrm{d}\vs
        - \int_{E_{i,j}} \nabla \cdot \matr{H}(\vs) \nabla u^{n+1}(\vs)\mathrm{d}\vs 
        + \int_{E_{i,j}} \nabla\cdot\vomega(\vs) u^{n+1}(\vs)\mathrm{d}\vs \right\} \\ 
        + \int_{E_{i,j}} \tau(B^{n+1}(\vs)-B^{n}(\vs))\mathrm{d}\vs,
\end{gather*}
The integrals on the first line and the first term on the second line is approximated by assuming that the integrands are constant in the grid cells. I.e.,
\begin{gather*}
    \int_{E_{i,j}} u^{n+1}(\vs)\mathrm{d}\vs \approx V_{i,j} u_{i,j}^{n+1}, \quad \int_{E_{i,j}} u^{n}(\vs)\mathrm{d}\vs \approx V_{i,j} u_{i,j}^{n} \\
    \text{and}\quad \int_{E_{i,j}} \kappa^2(\vs)u^{n+1}(\vs)\mathrm{d}\vs \approx V_{i,j} \kappa^2_{i,j} u_{i,j}^{n+1},
\end{gather*}
where $\kappa^2_{i,j} = \kappa(\vs_{i,j})^2$ and $u_{i,j}^{n+1} = u^{n+1}(\vs_{i,j})$. Similarily for the last term
\[
    \int_{E_{i,j}} \tau(B^{n+1}(\vs)-B^{n}(\vs))\mathrm{d}\vs \approx V_{i,j}\tau (B^{n+1}_{i,j}-B^n_{i,j}) \approx V_{i,j}\tau\sqrt{\Delta t} \phi_{i,j}^n,
\]
where $B^n_{i,j} = B^n(\vs_{i,j})$, and $\phi_{1,1}^n, \ldots, \phi_{M,N}^n$ arises from FVM applied to the purely spatial problem $\WM(\kappa_\mathrm{F}(\cdot), \matr{H}_\mathrm{F}(\cdot))$. We refer to \citet{fuglstad_exploring_2014,fuglstad_does_2015} for details on discretizing a spatial SPDE with FVM. However, the end result is that $\vphi^n = (\phi_{1,1}^n, \ldots, \phi_{M,N}^n) \sim \mathcal{N}(\boldsymbol{0}, \matr{Q}_\mathrm{F}^{-1})$ for a sparse matrix $\matr{Q}_\mathrm{F}$ and $\vphi^0, \ldots, \vphi^{T-2}$ are independent.

To handle the integrals of the advection and diffusion terms, we use the divergence theorem
\begin{equation*}
    \int_{E_{i,j}} \nabla \cdot \mathbf{F}(\vs)\enspace \mathrm{d}\vs  = \oint_{\partial E_{i,j}} \mathbf{F}(\vs) \cdot \boldsymbol{n}(\vs)\enspace\mathrm{d}A,
\end{equation*}
where $\mathbf{F}(\cdot)$ is a vector field, $\boldsymbol{n}(\cdot)$ is outward-facing normal vector to the respective faces, $\mathrm{d}A$ is the line element, and the line integral follows the positive orientation. This gives
\begin{align*}
        \int_{E_{i,j}} \nabla \cdot \matr{H}(\vs) \nabla u^{n+1}(\vs)\mathrm{d}\vs &= \oint_{\partial E_{i,j}} \nabla u^{n+1}(\vs)\cdot(\matr{H}(\vs)\boldsymbol{n}(\vs)) \dA, \\ 
        \int_{E_{i,j}} \nabla\cdot\vomega(\vs) u^{n+1}(\vs)\mathrm{d}\vs 
        &= \oint_{\partial E_{i,j}} u^{n+1}(\vs)(\vomega(\vs)\cdot\boldsymbol{n}(\vs))\dA.
\end{align*}

Thus using the approximations and the divergence theorem,
\begin{equation}
    \begin{aligned}
        \label{app:eq:advection_diffusion_fvm}
        V_{i,j}u_{i,j}^{n+1} &= V_{i,j}u_{i,j}^n - \dt V_{i,j} \kappa^2_{i,j}u_{i,j}^{n+1}  \\
        &\phantom{=}- \dt\oint_{\partial E_{i,j}} \left[-\nabla u^{n+1}(\vs) \cdot \{\matr{H}(\vs)\vn(\vs)\} + u^{n+1}(\vs)\{\vomega(\vs)\cdot\vn(\vs)\}\right]\dA \\
        &\phantom{=}+V_{i,j} \tau\sqrt{\dt} \phi_{i,j}^{n+1}. 
    \end{aligned}
\end{equation}
The second line is subtracting the net flow leaving the grid cell.
For approximating the flow arising from diffusion we refer to \citet{fuglstad_exploring_2014}, and we detail the approximation of the flow arising from advection in the next subsection. 

\subsection{Approximating flow arising from advection}
We consider a generic time step and drop superscripts to denote dependence on the time step, and decompose and approximate the net flow out of the domain from advection term in Equation~\eqref{app:eq:advection_diffusion_fvm} as 
\begin{equation*}
    \oint_{\partial E_{i,j}} u(\vs)(\vomega(\vs)\cdot\vn(\vs))\dA \approx \hat{W}_{i,j}^\rR + \hat{W}_{i,j}^\rL + \hat{W}_{i,j}^\rU + \hat{W}_{i,j}^\rD,
\end{equation*}
where the integral is split into integrals over each face that are approximated as
\begin{equation*}
    \begin{aligned}
         \int_{A_{i,j}^\text{Dir}} u(\vs)\left(\vomega(\vs)\cdot\vn(\vs)\right)\dA 
        \approx &   u(\vs_{i,j}^\text{Dir})\left(\vomega(\vs_{i,j}^\text{Dir})\cdot\vn(\vs_{i,j}^\text{Dir})\right)\int_{A_{i,j}^\text{Dir}} \dA \\
        = &  u(\vs_{i,j}^\text{Dir})\left(\vomega(\vs_{i,j}^\text{Dir})\cdot\vn(\vs_{i,j}^\text{Dir})\right)  |A_{i,j}^\text{Dir}| \\
        = &\tilde{W}_{i,j}^{\text{Dir}} .
    \end{aligned}
\end{equation*}
Here, $|A_{i,j}^\text{Dir}|$ is the length of the face $A_{i,j}^\text{Dir}$, and $\vs_{i,j}^\text{Dir}$ is the center of the face $A_{i,j}^\text{Dir}$,
i.e. if the right face is considered $|A_{i,j}^\rR| = h_y$ and $\vs_{i,j}^\rR = \vs_{i+1/2,j}$.

We use an upwind scheme \citep[Section 20.3]{eymard2000finite}, where we take the direction of the flow into account. When considering the flow over a given face, we only consider the upwind cell according to the advection vector field at the centre of the face. E.g., if we are considering the right face of a grid cell
and the flow goes to the left, we use the value of the field in the grid cell to the right.
This is known to help with numerical stability and is a common scheme in the finite volume method.

Let $\omega^x_{i\pm 1/2, j}$ denote the $x$-component of $\vomega(\vs_{i\pm 1/2, j})$ and $\omega^y_{i, j\pm 1/2}$ denote the $y$-component of $\vomega(\vs_{i, j \pm 1/2})$.
First, we consider the right face, and the upwind scheme gives
\begin{equation*}
\hat{W}_{i,j}^\rR = h_y\left[\frac{|\omega_{i+1/2,j}^{x}| + \omega_{i+1/2,j}^{x}}{2} u_{i,j}
- \frac{|\omega_{i+1/2,j}^{x}| - \omega_{i+1/2,j}^{x}}{2}u_{i+1,j}\right].
\end{equation*}
Similarly, we have for the left face
\begin{equation*}
\hat{W}_{i,j}^\rL = h_y\left[\frac{|\omega_{i-1/2,j}^{x}| - \omega_{i-1/2,j}^{x}}{2} u_{i,j}
- \frac{|\omega_{i-1/2,j}^{x}| + \omega_{i-1/2,j}^{x}}{2}u_{i-1,j}\right],
\end{equation*}
and for the top and bottom faces,
\begin{equation*}
\hat{W}_{i,j}^\rU = h_x\left[\frac{|\omega_{i,j+1/2}^{y}| + \omega_{i,j+1/2}^{y}}{2} u_{i,j}
- \frac{|\omega_{i,j+1/2}^{y}| - \omega_{i,j+1/2}^{y}}{2}u_{i,j+1}\right],
\end{equation*}
\begin{equation*}
\hat{W}_{i,j}^\rD = h_x\left[\frac{|\omega_{i,j-1/2}^{y}| - \omega_{i,j-1/2}^{y}}{2} u_{i,j}
- \frac{|\omega_{i,j-1/2}^{y}| + \omega_{i,j-1/2}^{y}}{2}u_{i,j-1}\right].
\end{equation*}

We vectorize the above expressions and define the vector of all spatial locations for timestep $n$ as
\[
\vu^{n}= (u_{0,0}^{n},u_{1,0}^{n},\ldots,u_{M-1,0}^{n},u_{0,1}^{n},u_{1,1}^{n},\ldots,u_{M-1,N-1}^{n})^\rT,
\]
for $n = 0, \ldots, T-1$. We introduce the advection matrix $\matr{A}_{\vomega}$ such that 
$\matr{A}_{\vomega}\vu^{n+1}$ gives the advection term in Equation~\eqref{app:eq:advection_diffusion_fvm} using the above approximations.
Thus $\matr{A}_{\vomega}$ is a $MN\times MN$ matrix,
where the local solution of the advection term for location $(i,j)$ is collected
in the $(jM +i)$th row of the matrix. In the following equations, we show the non-zero elements. For the diagonal elements we have
\begin{equation*}
\begin{aligned}
\left(\matr{A}_{\vomega}\right)_{jM + i,jM + i} = & \frac{h_y}{2}\left[|\omega_{i+1/2,j}^{x}| + \omega_{i+1/2,j}^{x} + |\omega_{i-1/2,j}^{x}| -
\omega_{i-1/2,j}^{x}\right]  + \\
& + \frac{h_x}{2}\left[|\omega_{i,j+1/2}^{y}| + \omega_{i,j+1/2}^{y} 
+ |\omega_{i,j-1/2}^{y}| - \omega_{i,j-1/2}^{y}\right].
\end{aligned}
\end{equation*}
Let $i_p = i + 1$, $i_n = i - 1$, $j_p = j + 1$, and $j_n = j - 1$, then the
$(jM+i_p)$th, $(jM+i_n)$th, $(j_pM+i)$th, and $(j_nM+i)$th elements of the $(jM+i)$th row are
\begin{equation*}
\left(\matr{A}_{\vomega}\right)_{jM + i,jM + i_p} = - \frac{h_y}{2}\left(|\omega_{i+1/2,j}^{x}| - \omega_{i+1/2,j}^{x}\right),
\end{equation*}
\begin{equation*}
\left(\matr{A}_{\vomega}\right)_{jM + i,jM + i_n} = - \frac{h_y}{2}\left(|\omega_{i-1/2,j}^{x}| + \omega_{i-1/2,j}^{x}\right),
\end{equation*}
\begin{equation*}
\left(\matr{A}_{\vomega}\right)_{jM + i,j_pM + i} = - \frac{h_x}{2}\left(|\omega_{i,j+1/2}^{y}| - \omega_{i,j+1/2}^{y}\right),
\end{equation*}
\begin{equation*}
\left(\matr{A}_{\vomega}\right)_{jM + i,j_nM + i} = - \frac{h_x}{2}\left(|\omega_{i,j-1/2}^{y}| + \omega_{i,j-1/2}^{y}\right).
\end{equation*}


\subsection{The fully descritized SPDE and the precision matrix}
Combining all previous steps, the fully discretized problem can be written in vector form as
\begin{equation*}
\left(\matr{D}_\rV  + \dt(\matr{D}_\rV\matr{D}_{\kappa^2} -
\matr{A}_\matr{H} + \matr{A}_{\vomega})\right)\vu^{n+1} =  \matr{D}_\rV\vu^{n} + \tau\sqrt{\dt}\matr{D}_\rV\vphi^{n},
\end{equation*}
where $\vphi^{n} = \matr{L}_\mathrm{F}^{-\rT}\vepsilon^{n}$ is the spatially smoothed innovations arising from
$\vepsilon^{0}, \ldots, \vepsilon^{n+1} \overset{\text{iid}}{\sim} \calN(0,\matr{I})$,
where $\matr{L}_\mathrm{F}$ is the Cholesky factor of $\mathbf{Q}_\mathrm{F}$, such that $\matr{Q}_\mathrm{F} = \matr{L}_\mathrm{F}\matr{L}_\mathrm{F}^\rT$.
We simplify the notation by introducing
\begin{equation*}
\begin{aligned}
\matr{A} = & \matr{D}_\rV  + \dt\left(\matr{D}_\rV\matr{D}_{\kappa^2} - \matr{A}_\matr{H} + \matr{A}_{\vomega}\right),\\
\matr{G} = & \matr{A}^{-1}\matr{D}_\rV,\\
\matr{E} = & \tau\sqrt{\dt} \matr{A}^{-1}\matr{D}_\rV\matr{L}_{\mathrm{F}}^{-\rT}.
\end{aligned}
\end{equation*}
Then the forward equations in time is
\begin{equation*}
\vu^{n+1} = \matr{G}\vu^{n} + \matr{E}\vepsilon^{n}, \quad n = 0, \ldots, T-2,
\end{equation*}
where $\vu^0\sim \mathcal{N}(\boldsymbol{0}, \matr{Q}_\mathrm{I}^{-1})$ arising from approximating $\WM(\kappa_\mathrm{I}(\cdot), \matr{H}_\mathrm{I}(\cdot))$.
However, note that computationally one would not compute the dense matrices $\matr{G}$ and $\matr{E}$ when using this equation for simulation.

Let us now collect all these timesteps into the vectors
\[
    \vu = \begin{bmatrix} \vu^0 \\ \vu^1 \\ \vdots \\ \vu^{T-1} \end{bmatrix}, \quad \text{and} \quad \vepsilon = \begin{bmatrix}\vepsilon^0 \\ \vdots \\ \vepsilon^{T-2}\end{bmatrix}.
\]
We then have
\begin{equation}
\label{app:eq:lin_rel}
\vu = \matr{R} \begin{bmatrix}
\vu^0 \\
\vepsilon
\end{bmatrix},
\end{equation}
where
$\matr{R}$ is a block matrix constructed from the matrices $\matr{G}$ and $\matr{E}$
as
\begin{equation}
\matr{R} = \begin{bmatrix}
\matr{I} & 0 &  0 & 0 & \dots & 0 & 0 \\
\matr{G} & \matr{E} & 0 & 0 &\dots & 0 & 0 \\
\matr{G}^2 & \matr{G}\matr{E} & \matr{E} & 0 & \dots & 0 & 0 \\
\matr{G}^3 & \matr{G}^2\matr{E} & \matr{G}\matr{E} & \matr{E} & \dots & 0 & 0 \\
\vdots & \vdots & \vdots & \vdots &\ddots & \vdots & \vdots \\
\matr{G}^{T-1} & \matr{G}^{T-2}\matr{E} &  \matr{G}^{T-3}\matr{E} & \matr{G}^{T-4}\matr{E} & \dots & \matr{G}\matr{E} & \matr{E}
\end{bmatrix}.
\end{equation}
We have 
\begin{equation*}
    \pi\left(\begin{bmatrix} \vu^0 \\ \vepsilon \end{bmatrix}\right) \propto \exp\left(-\frac{1}{2}
    \begin{bmatrix} (\vu^0)^\mathrm{T} & \vepsilon^\mathrm{T} \end{bmatrix}
    \begin{bmatrix}
        \matr{Q}_\mathrm{I}^{-1} & 0 & \dots & 0 \\ 0 & \matr{I} &\dots  & 0 \\
        \vdots & \vdots & \ddots & \vdots \\ 0 & 0 & \dots & \matr{I} \\
    \end{bmatrix}
    \begin{bmatrix} \vu^0 \\ \vepsilon \end{bmatrix}\right).
\end{equation*}
Equation~\eqref{app:eq:lin_rel} is a linear transformation and we can write the distribution of $\vu$ as
\begin{equation*}
\begin{aligned}
\pi(\vu)\propto &\exp\left(-\frac{1}{2}\vu^\rT\matr{Q}\vu\right) \\
 = & \exp\left(-\frac{1}{2}\vu^\rT\matr{R}^{-\rT}\begin{bmatrix}
\matr{Q}_\mathrm{I} & 0 & \dots & 0 \\ 0 & \matr{I} &\dots  & 0 \\
\vdots & \vdots & \ddots & \vdots \\ 0 & 0 & \dots & \matr{I} \\
\end{bmatrix}
\matr{R}^{-1}\vu\right).
\end{aligned}
\end{equation*}
Because of the block structure of $\matr{R}$, we can find its inverse as
\begin{equation*}
\matr{R}^{-1} = \begin{bmatrix}
\matr{I} & 0 &  0 & 0 & \dots & 0 & 0 \\
-\matr{E}^{-1}\matr{G} & \matr{E}^{-1} & 0 & 0 &\dots & 0 & 0 \\
0 & -\matr{E}^{-1}\matr{G} & \matr{E}^{-1} & 0 & \dots & 0 & 0 \\
0 & 0 & -\matr{E}^{-1}\matr{G} & \matr{E}^{-1} & \dots & 0 & 0 \\
\vdots & \vdots & \vdots & \vdots &\ddots & \matr{E}^{-1} & \vdots \\
0 & 0 & 0 & 0 & \dots & -\matr{E}^{-1}\matr{G} & \matr{E}^{-1}
\end{bmatrix},
\end{equation*}
where the inverse of $\matr{E}$ and the relationship $\matr{F} = \matr{E}\matr{E}^\rT$ is given by 
\begin{equation*}
\begin{aligned}
\matr{E}^{-1} = & \frac{1}{\tau\sqrt{\dt}} \matr{L}_\mathrm{F}^\mathrm{T}\matr{D}_V^{-1}\matr{A}\\
\matr{F}^{-1} = & \matr{E}^{-\rT}\matr{E}^{-1} = \frac{1}{\tau^2\dt} \matr{A}^\rT\matr{D}_V^{-1}\matr{Q}_S\matr{D}_V^{-1}\matr{A}.
\end{aligned}
\end{equation*}

Finally, we can express the global precision matrix as
\begin{equation*}
\matr{Q} = \begin{bmatrix}
\matr{Q}_\mathrm{I} + \matr{G}^\rT\matr{F}^{-1}\matr{G} & -\matr{G}^\rT\matr{F}^{-1} &  0 & \dots  & 0 \\
-\matr{F}^{-1}\matr{G} & \matr{F}^{-1} + \matr{G}^\rT\matr{F}^{-1}\matr{G} & -\matr{G}^\rT\matr{F}^{-1} &\ddots &0 \\
\vdots & \ddots & \ddots &\ddots & \vdots  \\
\vdots & \ddots &-\matr{F}^{-1}\matr{G} & \matr{F}^{-1} + \matr{G}^\rT\matr{F}^{-1}\matr{G} & -\matr{G}^\rT\matr{F}^{-1}\\ 
0 & \dots & 0 & -\matr{F}^{-1}\matr{G} &  \matr{F}^{-1}
\end{bmatrix}.
\end{equation*}

\section{Code implementations}
\label{app:sec:impl}

The models described in this work are implemented in a Python package 
named \texttt{spdepy}. The package is available at \url{https://github.com/berild/spdepy}
and is under development. Currently, the package is not available on PyPI, but
can be installed using poetry.

The package is built from scratch using mostly NumPy and Cython for C++ interaction.
The construction of the precision matrix components is done in C++ in order to speed up the
computations. The models are only available for 2D spatial fields with a time component, 
but an extension to 3D and time is possible.

Available SPDEs in the package include the advection-diffusion SPDE, the separable space-time model, and the Whittle-Matérn SPDE. Users have the flexibility to specify whether the parameters are spatially varying or constant, allowing for various combinations. For instance, spatial models can be stationary or non-stationary, as well as isotropic or anisotropic. Additionally, spatio-temporal models can incorporate constant or spatially varying advection and diffusion, or any combination of these factors. It is also possible to treat advection as a covariate. Users can input any initial condition, provided it aligns with one of the aforementioned models included in the package.

The user can choose which optimization algorithm to use; currently, the package
supports gradient descent, Adadelta, Adagrad, Rmsprop, and Adam. All of which are
stochastic as we use the Hutchinson estimator in the gradient calculations.
Also implemented is the AdaHessian optimizer, but is unavailable in the current version.

Lastly, also included in the package is the ocean dataset generated from the 
numerical model SINMOD. It is split into three datasets, one for training,
one for validation, and one for testing. The testing dataset consists of real
measurements collected with an AUV that is not used in this work. 

\newpage
\bibliographystyle{apalike}
\bibliography{ref}

\begin{thebibliography}{}

\bibitem[Adler and Taylor, 2007]{adler2009random}
Adler, R.~J. and Taylor, J.~E. (2007).
\newblock {\em Random fields and geometry}.
\newblock Springer Science \& Business Media, NY: New York.

\bibitem[Banerjee et~al., 2003]{banerjee_hierarchical_2003}
Banerjee, S., Carlin, B.~P., and Gelfand, A.~E. (2003).
\newblock {\em Hierarchical {Modeling} and {Analysis} for {Spatial} {Data}}.
\newblock CRC Press.

\bibitem[Banerjee et~al., 2008]{banerjee_gaussian_2008}
Banerjee, S., Gelfand, A.~E., Finley, A.~O., and Sang, H. (2008).
\newblock Gaussian {Predictive} {Process} {Models} for {Large} {Spatial} {Data} {Sets}.
\newblock {\em Journal of the Royal Statistical Society. Series B (Statistical Methodology)}, 70(4):825--848.

\bibitem[Berild and Fuglstad, 2023]{berild_spatially_2023}
Berild, M.~O. and Fuglstad, G.-A. (2023).
\newblock Spatially varying anisotropy for gaussian random fields in three-dimensional space.
\newblock {\em Spatial Statistics}, 55:100750.

\bibitem[Berild et~al., 2024]{berild_efficient_2024}
Berild, M.~O., Ge, Y., Eidsvik, J., Fuglstad, G.-A., and Ellingsen, I. (2024).
\newblock Efficient {3D} real-time adaptive {AUV} sampling of a river plume front.
\newblock {\em Frontiers in Marine Science}, 10.

\bibitem[Bolin and Kirchner, 2020]{bolin_rational_2020}
Bolin, D. and Kirchner, K. (2020).
\newblock The {Rational} {SPDE} {Approach} for {Gaussian} {Random} {Fields} {With} {General} {Smoothness}.
\newblock {\em Journal of Computational and Graphical Statistics}, 29(2):274--285.

\bibitem[Butcher, 2016]{butcher2016numerical}
Butcher, J.~C. (2016).
\newblock {\em Numerical methods for ordinary differential equations}.
\newblock John Wiley \& Sons.

\bibitem[Cameletti et~al., 2013]{cameletti_spatio-temporal_2013}
Cameletti, M., Lindgren, F., Simpson, D., and Rue, H. (2013).
\newblock Spatio-temporal modeling of particulate matter concentration through the {SPDE} approach.
\newblock {\em AStA Advances in Statistical Analysis}, 97(2):109--131.

\bibitem[Carrizo~Vergara et~al., 2022]{carrizo_vergara_general_2022}
Carrizo~Vergara, R., Allard, D., and Desassis, N. (2022).
\newblock A general framework for {SPDE}-based stationary random fields.
\newblock {\em Bernoulli}, 28.

\bibitem[Clarotto et~al., 2024]{clarotto_spde_2023}
Clarotto, L., Allard, D., Romary, T., and Desassis, N. (2024).
\newblock The spde approach for spatio-temporal datasets with advection and diffusion.
\newblock {\em Spatial Statistics}, 62:100847.

\bibitem[Cressie and Johannesson, 2008]{cressie_fixed_2008}
Cressie, N. and Johannesson, G. (2008).
\newblock Fixed rank kriging for very large spatial data sets.
\newblock {\em Journal of the Royal Statistical Society: Series B (Statistical Methodology)}, 70(1):209--226.

\bibitem[Cressie and Wikle, 2011]{cressie_statistics_2011}
Cressie, N. and Wikle, C.~K. (2011).
\newblock {\em Statistics for spatio-temporal data}.
\newblock John Wiley \& Sons.

\bibitem[Da~Prato and Zabczyk, 2014]{da2014stochastic}
Da~Prato, G. and Zabczyk, J. (2014).
\newblock {\em Stochastic equations in infinite dimensions}.
\newblock Cambridge university press.

\bibitem[Eidsvik et~al., 2014]{eidsvik_estimation_2014}
Eidsvik, J., Shaby, B.~A., Reich, B.~J., Wheeler, M., and Niemi, J. (2014).
\newblock Estimation and {Prediction} in {Spatial} {Models} {With} {Block} {Composite} {Likelihoods}.
\newblock {\em Journal of Computational and Graphical Statistics}, 23(2):295--315.

\bibitem[Eymard et~al., 2000]{eymard2000finite}
Eymard, R., Gallou{\"e}t, T., and Herbin, R. (2000).
\newblock Finite volume methods.
\newblock {\em Handbook of numerical analysis}, 7:713--1018.

\bibitem[Foss et~al., 2022]{foss_using_2022}
Foss, K.~H., Berget, G.~E., and Eidsvik, J. (2022).
\newblock Using an autonomous underwater vehicle with onboard stochastic advection-diffusion models to map excursion sets of environmental variables.
\newblock {\em Environmetrics}, 33(1):e2702.

\bibitem[Fossum et~al., 2021]{fossum2021learning}
Fossum, T.~O., Travelletti, C., Eidsvik, J., Ginsbourger, D., and Rajan, K. (2021).
\newblock {Learning excursion sets of vector-valued Gaussian random fields for autonomous ocean sampling}.
\newblock {\em The Annals of Applied Statistics}, 15(2):597 -- 618.

\bibitem[Fuglstad and Castruccio, 2020]{fuglstad_compression_2020}
Fuglstad, G.-A. and Castruccio, S. (2020).
\newblock Compression of climate simulations with a nonstationary global spatiotemporal spde model.
\newblock {\em The Annals of Applied Statistics}, 14(2):542--559.

\bibitem[Fuglstad et~al., 2015a]{fuglstad_exploring_2014}
Fuglstad, G.-A., Lindgren, F., Simpson, D., and Rue, H. (2015a).
\newblock Exploring a new class of non-stationary spatial gaussian random fields with varying local anisotropy.
\newblock {\em Statistica Sinica}, pages 115--133.

\bibitem[Fuglstad et~al., 2015b]{fuglstad_does_2015}
Fuglstad, G.-A., Simpson, D., Lindgren, F., and Rue, H. (2015b).
\newblock Does non-stationary spatial data always require non-stationary random fields?
\newblock {\em Spatial Statistics}, 14:505--531.

\bibitem[Fuglstad et~al., 2019]{fuglstad2019constructing}
Fuglstad, G.-A., Simpson, D., Lindgren, F., and Rue, H. (2019).
\newblock Constructing priors that penalize the complexity of gaussian random fields.
\newblock {\em Journal of the American Statistical Association}, 114(525):445--452.

\bibitem[Ge et~al., 2023]{yaolin2023}
Ge, Y., Eidsvik, J., and Mo-Bjørkelund, T. (2023).
\newblock 3d adaptive auv sampling for classification of water masses.
\newblock {\em IEEE Journal of Ocean Engineering}, 48:626--639.

\bibitem[Gneiting and Raftery, 2007]{gneiting_strictly_2007}
Gneiting, T. and Raftery, A.~E. (2007).
\newblock Strictly {Proper} {Scoring} {Rules}, {Prediction}, and {Estimation}.
\newblock {\em Journal of the American Statistical Association}, 102(477).

\bibitem[Heaton et~al., 2019]{heatonEtAl2019}
Heaton, M.~J., Datta, A., Finley, A.~O., Furrer, R., Guinness, J., Guhaniyogi, R., Gerber, F., Gramacy, R.~B., Hammerling, D., Katzfuss, M., Lindgren, F., Nychka, D.~W., Sun, F., and Zammit-Mangion, A. (2019).
\newblock A case study competition among methods for analyzing large spatial data.
\newblock {\em Journal of Agricultural, Biological and Environmental Statistics}, 24:398--425.

\bibitem[Heine, 1955]{heine_models_1955}
Heine, V. (1955).
\newblock Models for two-dimensional stationary stochastic processes.
\newblock {\em Biometrika}, 42(1-2):170--178.

\bibitem[Hildeman et~al., 2021]{hildeman2021deformed}
Hildeman, A., Bolin, D., and Rychlik, I. (2021).
\newblock Deformed spde models with an application to spatial modeling of significant wave height.
\newblock {\em Spatial Statistics}, 42:100449.

\bibitem[Hutchinson, 1990]{hutchinson_stochastic_1990}
Hutchinson, M. (1990).
\newblock A stochastic estimator of the trace of the influence matrix for {Laplacian} smoothing splines.
\newblock {\em Communication in Statistics- Simulation and Computation}, 19(2):433--450.

\bibitem[Jones and Zhang, 1997]{jones_models_1997}
Jones, R.~H. and Zhang, Y. (1997).
\newblock Models for {Continuous} {Stationary} {Space}-{Time} {Processes}.
\newblock In Gregoire, T.~G., Brillinger, D.~R., Diggle, P.~J., Russek-Cohen, E., Warren, W.~G., and Wolfinger, R.~D., editors, {\em Modelling {Longitudinal} and {Spatially} {Correlated} {Data}}, pages 289--298, New York, NY. Springer.

\bibitem[Katzfuss and Guinness, 2021]{katzfuss2021general}
Katzfuss, M. and Guinness, J. (2021).
\newblock A {General} {Framework} for {Vecchia} {Approximations} of {Gaussian} {Processes}.
\newblock {\em Statistical Science}, 36(1):124--141.

\bibitem[Kingma and Ba, 2017]{kingma_adam_2017}
Kingma, D.~P. and Ba, J. (2017).
\newblock Adam: {A} {Method} for {Stochastic} {Optimization}.
\newblock {\em arXiv:1412.6980 [cs.LG]}.

\bibitem[LeVeque, 2002]{leveque_finite_2002}
LeVeque, R.~J. (2002).
\newblock {\em Finite {Volume} {Methods} for {Hyperbolic} {Problems}}.
\newblock Cambridge {Texts} in {Applied} {Mathematics}. Cambridge University Press, Cambridge.

\bibitem[Lindgren et~al., 2023]{lindgren_diffusion-based_2023}
Lindgren, F., Bakka, H., Bolin, D., Krainski, E., and Rue, H. (2023).
\newblock A diffusion-based spatio-temporal extension of {Gaussian} {Mat}{\textbackslash}'ern fields.
\newblock arXiv:2006.04917 [stat].

\bibitem[Lindgren et~al., 2022]{lindgren_spde_2022}
Lindgren, F., Bolin, D., and Rue, H. (2022).
\newblock The {SPDE} approach for {Gaussian} and non-{Gaussian} fields: 10 years and still running.
\newblock {\em Spatial Statistics}, 50:100599.

\bibitem[Lindgren et~al., 2011]{lindgren_explicit_2011}
Lindgren, F., Rue, H., and Lindström, J. (2011).
\newblock An explicit link between {Gaussian} fields and {Gaussian} {Markov} random fields: the stochastic partial differential equation approach.
\newblock {\em Journal of the Royal Statistical Society: Series B (Statistical Methodology)}, 73(4):423--498.

\bibitem[Liu et~al., 2022]{liu_statistical_2022}
Liu, X., Yeo, K., and Lu, S. (2022).
\newblock Statistical {Modeling} for {Spatio}-{Temporal} {Data} {From} {Stochastic} {Convection}-{Diffusion} {Processes}.
\newblock {\em Journal of the American Statistical Association}, 117(539):1482--1499.

\bibitem[Llamazares-Elias et~al., 2024]{llamazares2024parameterization}
Llamazares-Elias, L., Latz, J., and Lindgren, F. (2024).
\newblock A parameterization of anisotropic gaussian fields with penalized complexity priors.
\newblock {\em arXiv preprint arXiv:2409.02331}.

\bibitem[Patrizi et~al., 2020]{patrizi_adaptive_2020}
Patrizi, F., Manni, C., Pelosi, F., and Speleers, H. (2020).
\newblock Adaptive refinement with locally linearly independent {LR} {B}-splines: {Theory} and applications.
\newblock {\em Computer Methods in Applied Mechanics and Engineering}, 369:113230.

\bibitem[Pereira et~al., 2022]{pereira_geostatistics_2022}
Pereira, M., Desassis, N., and Allard, D. (2022).
\newblock Geostatistics for {Large} {Datasets} on {Riemannian} {Manifolds}: {A} {Matrix}-{Free} {Approach}.
\newblock {\em Journal of Data Science}, 20(4):512--532.

\bibitem[Porcu et~al., 2021]{porcu_30_2021}
Porcu, E., Furrer, R., and Nychka, D. (2021).
\newblock 30 {Years} of space–time covariance functions.
\newblock {\em WIREs Computational Statistics}, 13(2):e1512.

\bibitem[Rodríguez-Iturbe and Mejía, 1974]{rodriguez-iturbe_design_1974}
Rodríguez-Iturbe, I. and Mejía, J.~M. (1974).
\newblock The design of rainfall networks in time and space.
\newblock {\em Water Resources Research}, 10(4):713--728.

\bibitem[Ruder, 2017]{ruder_overview_2017}
Ruder, S. (2017).
\newblock An overview of gradient descent optimization algorithms.
\newblock {\em arXiv:1609.04747 [cs.LG]}.

\bibitem[Rue and Held, 2005]{rue_gaussian_2005}
Rue, H. and Held, L. (2005).
\newblock {\em Gaussian {Markov} random fields: theory and applications}, volume 104 of {\em Monographs on statistics and applied probability}.
\newblock Chapman \& Hall/CRC, Boca Raton, Fla.

\bibitem[Rue and Held, 2010]{col27}
Rue, H. and Held, L. (2010).
\newblock Markov random fields.
\newblock In Gelfand, A., Diggle, P., Fuentes, M., and Guttorp, P., editors, {\em Handbook of Spatial Statistics}, pages 171--200. CRC/Chapman \& Hall, Boca Raton, FL.

\bibitem[Sigrist et~al., 2014]{sigrist_stochastic_2014}
Sigrist, F., Künsch, H., and Stahel, W. (2014).
\newblock Stochastic {Partial} {Differential} {Equation} {Based} {Modelling} of {Large} {Space}–{Time} {Data} {Sets}.
\newblock {\em Journal of the Royal Statistical Society: Series B (Statistical Methodology)}, 77.

\bibitem[Simpson et~al., 2012]{simpson_order_2012}
Simpson, D., Lindgren, F., and Rue, H. (2012).
\newblock In order to make spatial statistics computationally feasible, we need to forget about the covariance function.
\newblock {\em Environmetrics}, 23(1):65--74.

\bibitem[Slagstad and McClimans, 2005]{slagstad_modeling_2005}
Slagstad, D. and McClimans, T.~A. (2005).
\newblock Modeling the ecosystem dynamics of the {Barents} sea including the marginal ice zone: {I}. {Physical} and chemical oceanography.
\newblock {\em Journal of Marine Systems}, 58(1):1--18.

\bibitem[Vecchia, 1988]{vecchia_estimation_1988}
Vecchia, A.~V. (1988).
\newblock Estimation and {Model} {Identification} for {Continuous} {Spatial} {Processes}.
\newblock {\em Journal of the Royal Statistical Society: Series B (Methodological)}, 50(2):297--312.

\bibitem[Wikle, 2010]{wikle_low-rank_2010}
Wikle, C.~K. (2010).
\newblock Low-{Rank} {Representations} for {Spatial} {Processes}.
\newblock In Gelfand, A.~E., Diggle, P., Guttorp, P., and Fuentes, M., editors, {\em Handbook of {Spatial} {Statistics}}, chapter~8, pages 107--117. CRC Press.

\bibitem[Yao et~al., 2021]{yao_adahessian_2021}
Yao, Z., Gholami, A., Shen, S., Mustafa, M., Keutzer, K., and Mahoney, M.~W. (2021).
\newblock {ADAHESSIAN}: {An} {Adaptive} {Second} {Order} {Optimizer} for {Machine} {Learning}.
\newblock arXiv:2006.00719 [cs, math, stat].

\end{thebibliography}

\end{document}